\documentclass[12pt]{iopart}
\usepackage{iopams,mathptm,times,graphicx}

\makeatletter

\newcounter{theorem}
\@addtoreset{theorem}{section}
\renewcommand\thetheorem{\arabic{section}.\arabic{theorem}}

\newenvironment{lemma}{\par\medskip\noindent\begingroup{\bf Lemma
             \stepcounter{theorem}\thetheorem.}\ \itshape
             \def\@currentlabel{\thetheorem}}{\endgroup\par\medskip}
\newenvironment{theorem}{\par\medskip\noindent\begingroup{\bf Theorem
             \stepcounter{theorem}\thetheorem.}\ \itshape
             \def\@currentlabel{\thetheorem}}{\endgroup\par\medskip}

\newenvironment{corollary}{\par\medskip\noindent\begingroup{\bf Corollary
             \stepcounter{theorem}\thetheorem.}\ \itshape
             \def\@currentlabel{\thetheorem}}{\endgroup\par\medskip}
\newenvironment{proof}{\par\noindent{\bf Proof.} }{\proofbox\par\medskip}

\def\proofbox{\hfill{\ensuremath\Box}}

\newdimen\LENB \newdimen\LENW \newdimen\THI
\newdimen\LENWH \newdimen\LENTOT \newcount\N
\def\vbrknlnele#1#2#3{
  \LENB=#1pt \LENW=#2pt \THI=#3pt
  \LENWH=\LENW \divide\LENWH by 2
  \LENTOT=\LENB \advance\LENTOT by \LENW
  \vbox to \LENTOT{
    \vbox to \LENWH{}
    \nointerlineskip
    \vbox to \LENB{\hbox to \THI{\vrule width \THI height \LENB}}
    \nointerlineskip
    \vbox to \LENWH{}
  }}

\def\vbrknln#1{
  \N=#1
  \vcenter{
    \vbox{
      \loop\ifnum\N>0
        \vbox to 4pt{\vbrknlnele{2}{2}{0.1}}
        \nointerlineskip
        \advance\N by -1
      \repeat
  }}}

\def\vbl#1{\hskip-5pt \vbrknln{#1} \hskip-5pt}

\def\hbrknlnele#1#2#3{
  \LENB=#1pt \LENW=#2pt \THI=#3pt
  \LENTOT=\LENB \advance\LENTOT by \LENW
  \vcenter{
    \vbox to \THI{
      \hbox to \LENTOT{
        \hfil
        \vrule width \LENB height \THI
        \hfil}
  }}}

\def\hblele{\hbrknlnele{2}{2.2}{0.1}}

\def\hblfil{\cleaders\hbox{$\mkern1mu \hblele \mkern1mu$}\hfill}

\makeatletter

\usepackage{ulem}

\def\journal#1&#2,{\begingroup \let\journal=\dummyjournal
               \it #1\unskip~\bf\ignorespaces #2\rm,\endgroup}
\def\dummyjournal{\errmessage{Reference foul up: nested \journal macros}}

\eqnobysec

\def\eqref#1{(\ref{#1})}
\begin{document}
\title[An integrable semi-discretization of the Camassa-Holm equation]
  {An integrable semi-discretization of the Camassa-Holm equation and its
  determinant solution}
\author{Yasuhiro Ohta$^{1}$, Ken-ichi Maruno$^2
\footnote{e-mail: kmaruno@utpa.edu}$ and Bao-Feng Feng$^2$
}
\address{$^1$~Department of Mathematics,
Kobe University, Rokko, Kobe 657-8501, Japan
}
\address{$^2$~Department of Mathematics,
The University of Texas-Pan American,
Edinburg, TX 78541}
\date{\today}
\def\submitto#1{\vspace{28pt plus 10pt minus 18pt}
     \noindent{\small\rm To be submitted to : {\it #1}\par}}

\begin{abstract}
An integrable semi-discretization of
the Camassa-Holm equation is presented.
The keys of its construction are bilinear forms and
determinant structure of solutions of the CH equation.
Determinant formulas of $N$-soliton solutions of the continuous and
semi-discrete Camassa-Holm equations are presented.
Based on determinant formulas, we can generate multi-soliton,
multi-cuspon and multi-soliton-cuspon solutions.
Numerical computations using the integrable
semi-discrete Camassa-Holm equation are performed.
It is shown that the integrable semi-discrete Camassa-Holm
equation gives very accurate numerical
results even in the cases of cuspon-cuspon and soliton-cuspon
interactions.
The numerical computation for an initial
value condition, which is not an exact solution, is also presented.
\par
\kern\bigskipamount\noindent
\today
\end{abstract}

\kern-\bigskipamount

\submitto{\JPA}

\section{Introduction}
The  Camassa-Holm (CH) equation
\begin{equation}
w_T+2\kappa^2w_X-w_{TXX}+3ww_X=2w_Xw_{XX}+ww_{XXX}.\label{CH-eq}
\end{equation}
has attracted considerable interest since it has been derived as a model
equation for shallow-water waves \cite{CH}.
Here, $w=w(X,T)$, $\kappa$ is a positive parameter
and the subscripts $T$ and $X$ appended to $w(X,T)$
denote partial differentiation.
Originally, this equation has been found in a mathematical
search of recursion operators connected with the integrable
partial differential equations \cite{FF}.
The CH equation has been
shown to be completely integrable.
In the case of $\kappa=0$,
the CH equation admits peakon solutions which are represented by
piecewise analytic functions \cite{CHH}.
Schiff obtained single- and two-soliton solutions in a parametric form
by using the B\"acklund transformation \cite{Schiff}.
An approach based on the
inverse scattering transform method (IST) provides
an explicit form of the inverse mapping
in terms of Wronskian \cite{Constantin,Johnson,Li,Li2}.
The $N$-soliton solution was also constructed by using the
Hirota bilinear method \cite{Parker,Parker2,Parker3,Matsuno}
(See also \cite{Hone}).
The key point of computations of the $N$-soliton solution
by the Hirota bilinear method in those papers
is the relationship between the CH equation and the AKNS shallow
water wave equation \cite{AKNS}.
When $\kappa\not=0$, cusped solitary wave solutions,
as well as analytic soliton solutions, were found in
\cite{Olver,Kraenkel,Kraenkel2}.
In \cite{DaiLi,Parker-cusp}, the interaction of cusped soliton
(cuspon) was studied in detail.

It is extremely difficult to perform numerical computations of
the CH equation due to the singularities of cuspon and peakon
solutions.
So far, several numerical computations
of the CH equation were presented
\cite{Kalisch,Lee,Holden1,Holden2,Artebrant,Coclite,Cohen}.
However, none of these numerical methods
gives satisfactory results for soliton-cuspon
and cuspon-cuspon interactions.

Integrable discretizations of soliton equations
have received considerable attention recently \cite{SIDE,DIS}.
Ablowitz and Ladik proposed
how to construct
integrable discrete analogues of soliton equations based on
Lax pairs \cite{AL1,AL2}.
Hirota proposed another method to construct
integrable discrete analogues of
soliton equations
based on bilinear equations \cite{Hirota-d1,Hirota-d2,Hirota-d3}.
Applications of integrable discretizations of soliton equations were
considered in various fields \cite{Nishinari,Nakamura1,Nakamura2}.

The purpose in the present paper
is to present an integrable semi-discretization
of the CH equation through a bilinear approach
and to perform numerical computations by using integrable
scheme. We show that the integrable
scheme gives very accurate numerical results even for the
soliton-solition and cuspon-cuspon interactions.

The outline of this paper is as follows.
In \S 2, we give bilinear forms and a determinant formula of the
$N$-soliton
solution of equation (\ref{CH-eq}). As far as we are concerned,
bilinear forms directly related to the CH equation
have not been known yet.
These bilinear forms can help us to understand mathematical structure
of the CH equation more thoroughly.
Based on these bilinear equations, we obtain the determinant
formula for the CH equation by using determinant technique
\cite{OHTI,OKMS,FN,FN2}.
In \S 3, we give an integrable semi-discrete Camassa-Holm equation and its
determinant formula of
the $N$-soliton solution.
In \S 4, we present results of numerical computations by using the
proposed
integrable semi-discrete CH equation.

\section{Bilinear equations and determinant formulas of the
Camassa-Holm equation}
\noindent
In this section, we give bilinear equations and the $N$-soliton
solution of the CH equation.
All of the existing works in constructing $N$-soliton solutions of
the CH equation using Hirota bilinear method take advantage of the
relationship between the CH equation and the AKNS shallow water wave
equation \cite{Parker,Parker2,Parker3,Matsuno,AKNS}. So far, the
bilinear forms which are directly related to the CH equation remain
unknown.
%%%The past works of the CH equation using Hirota bilinear method used the
%%%relationship with the AKNS shallow water wave equation and
%%%construct solutions using bilinear equations of the AKNS shallow water
%%%wave equation \cite{Parker,Parker2,Parker3,Matsuno,AKNS}. Thus they didn't
%%%show bilinear equations which are directly obtained from the CH
%%%equation.
In the present paper,
we give bilinear equations obtained from the CH equation directly
and derive
the determinant solution by using the determinant technique.
Our formulation in this section is crucial for
the derivation of
integrable semi-discretization
of the CH equation.

\begin{lemma}
Bilinear equations
\begin{equation}
\left\{\begin{array}{l}\displaystyle
\left(\frac{1}{2}D_tD_x-1\right)f\cdot f=-gh\,,
\\[5pt]\displaystyle
\frac{1}{2}D_t(D_y-2cD_x)f\cdot f=-D_xg\cdot h\,,
\\[5pt]\displaystyle
\frac{1}{2}D_sD_xf\cdot f=D_tg\cdot h\,,
\\[5pt]\displaystyle
\left(\frac{1}{2}D_s(D_y-2cD_x)-2\right)f\cdot f=(D_tD_x-2)g\cdot h\,,
\end{array}\right.\label{2dbilinear}
\end{equation}
have a determinant solution
$$
f=\tau_0\,,
\qquad
g=\tau_1\,,
\qquad
h=\tau_{-1}\,,
$$
$$
\tau_n=\left|\matrix{
\psi_1^{(n)} &\psi_1^{(n+1)} &\cdots &\psi_1^{(n+N-1)} \cr
\psi_2^{(n)} &\psi_2^{(n+1)} &\cdots &\psi_2^{(n+N-1)} \cr
\vdots       &\vdots         &       &\vdots           \cr
\psi_N^{(n)} &\psi_N^{(n+1)} &\cdots &\psi_N^{(n+N-1)}}\right|\,,
$$
where
$$
\psi_i^{(n)}=a_{i,1}(p_i-c)^ne^{\xi_i}+a_{i,2}(q_i-c)^ne^{\eta_i}\,,
$$
$$
\xi_i=p_ix+p_i^2y+\frac{1}{p_i-c}t+\frac{1}{(p_i-c)^2}s+\xi_{i0}\,,
$$
$$
\eta_i=q_ix+q_i^2y+\frac{1}{q_i-c}t+\frac{1}{(q_i-c)^2}s+\eta_{i0}\,.
$$
\end{lemma}
\begin{proof}
Consider the following Casorati determinant solution,
\begin{eqnarray}
 \tau_n = \left|\begin{array}{cccc}
  \psi_1^{(n)} &\psi_1^{(n+1)} &\cdots
   &\psi_1^{(n+N-1)} \cr
  \psi_2^{(n)} &\psi_2^{(n+1)} &\cdots
   &\psi_2^{(n+N-1)} \cr
  \vdots               &\vdots                 &
   &\vdots                 \cr
  \psi_N^{(n)} &\psi_N^{(n+1)} &\cdots
   &\psi_N^{(n+N-1)}\cr
\end{array}
 \right|,\label{determinant-bt}
\end{eqnarray}
where $\psi_i^{(n)}$'s are arbitrary functions of four continuous
independent variables, $x$, $y$, $t$ and $s$, which
satisfy linear dispersion relations,
\begin{eqnarray}
&&\partial_x\psi_i^{(n)} = \psi_i^{(n+1)}+c\psi_i^{(n)}\,,
\label{x-dispersion}\\
&&\partial_y\psi_i^{(n)} = \partial_x^2\psi_i^{(n)}\,,\nonumber\\
&&\qquad \, \, =\psi_i^{(n+2)}+2c\psi_i^{(n+1)}+c^2\psi_i^{(n)}\,,
\label{y-dispersion}\\
&&
\partial_t\psi_i^{(n)} = \psi_i^{(n-1)}\,,
\label{t-linear}\\
&&
\partial_s\psi_i^{(n)} = \psi_i^{(n-2)}\,.
\label{s-linear}
\end{eqnarray}
Thus we can choose $\psi_i^{(n)}$ as follows:
$$
\psi_i^{(n)}=a_{i,1}(p_i-c)^ne^{\xi_i}+a_{i,2}(q_i-c)^ne^{\eta_i}\,,
$$
$$
\xi_i=p_ix+p_i^2y+\frac{1}{p_i-c}t+\frac{1}{(p_i-c)^2}s+\xi_{i0}\,,
$$
$$
\eta_i=q_ix+q_i^2y+\frac{1}{q_i-c}t+\frac{1}{(q_i-c)^2}s+\eta_{i0}\,.
$$
For simplicity, we introduce a convenient notation,
\begin{eqnarray}
\fl |{n_1},
  {n_2},
  \cdots,
  {n_N}|
 = \left|\matrix{
  \psi_1^{(n_1)} &\psi_1^{(n_2)}&\cdots
   &\psi_1^{(n_N)}\cr
  \psi_2^{(n_1)} &\psi_2^{(n_2)}&\cdots
   &\psi_2^{(n_N)} \cr
  \vdots                     &\vdots                     &
   &\vdots                     \cr
  \psi_N^{(n_1)} &\psi_N^{(n_2)} &\cdots
   &\psi_N^{(n_N)} \cr}
 \right|.
\end{eqnarray}
In this notation, the solution for the above bilinear forms, $\tau_n$,
is rewritten as
\begin{eqnarray}
 \tau_n=|n, n+1,\cdots,n+N-1|\,.\label{n-tau}
\end{eqnarray}
 We show that the above $\tau_n$ actually satisfies
the bilinear equations (\ref{2dbilinear}) by using the Laplace expansion
technique \cite{FN,FN2}.

The differential formulas for $\tau$ are given by
\begin{eqnarray}
&&\fl (\partial_x-Nc)\tau_n=|n,n+1,\cdots,n+N-2,n+N|\,, \label{x-dif}
\end{eqnarray}
\begin{eqnarray}
&&\fl (\partial_y-2c\partial_x+Nc^2)\tau_n=|n,n+1,\cdots,n+N-2,n+N+1|
\nonumber\\
&&\fl \quad -|n,n+1,\cdots,n+N-3,n+N-1,n+N|\,, \label{y-dif}
\end{eqnarray}
\begin{eqnarray}
&&\fl \partial_t\tau_n=|n-1,n+1,\cdots,n+N-1|\,, \label{t-dif}
\end{eqnarray}
\begin{eqnarray}
&&\fl \partial_s\tau_{n}=|n-2,n+1,\cdots, n+N-1|
-|n-1,n,n+2,\cdots,n+N-1|\,, \label{s-dif}
\end{eqnarray}
\begin{eqnarray}
&&\fl (\partial_t(\partial_x-Nc)-1)\tau_n=|n-1,n+1,\cdots,n+N-2,n+N|
\,, \label{xt-dif}
\end{eqnarray}
\begin{eqnarray}
&&\fl \partial_t(\partial_y-2c\partial_x+Nc^2)\tau_n
=|n-1,n+1,\cdots,n+N-2,n+N+1|\nonumber\\
&&\fl \quad -|n-1,n+1,\cdots,n+N-3,n+N-1,n+N|\,, \label{yt-dif}
\end{eqnarray}
\begin{eqnarray}
&&\fl \partial_s(\partial_x-Nc)\tau_n=|n-2,n+1,\cdots,n+N-2,n+N|\nonumber\\
&&\fl \quad -|n-1,n,n+2,\cdots,n+N-2,n+N|\,, \label{xs-dif}
\end{eqnarray}
\begin{eqnarray}
&&\fl (\partial_s(\partial_y-2c\partial_x+Nc^2)-2)\tau_n
=|n-2,n+1,\cdots,n+N-2,n+N+1|\nonumber\\
&&\fl \quad -|n-1,n,n+2,\cdots,n+N-2,n+N+1|
-|n-2,n+1,\cdots,n+N-3,n+N-1,n+N|\nonumber\\
&&\fl \quad +|n-1,n,n+2,\cdots,n+N-3,n+N-1,n+N|\,, \label{ys-dif}
\end{eqnarray}
which are proved by using the linear dispersion relations
(\ref{x-dispersion})-(\ref{s-linear}). (See Appendix) \\
\quad \\
{\bf The first equation of eqs.~(\ref{2dbilinear})}\\
Let us introduce an identity for $2N\times 2N$ determinant,
$$
\left|\matrix{
n-1 &\vbl4 & n+1 & \cdots &n+N-2 &\vbl4 &n+N &\vbl4 &n
&\vbl4
& &\hbox{\O} & &\vbl4 &n+N-1 \cr
\multispan{15}\hblfil \cr
n-1 &\vbl4 & & \hbox{\O}    &    &\vbl4 &n+N &\vbl4 & n &\vbl4
&n+1 &\cdots &n+N-2 &\vbl4 &n+N-1\cr}
 \right| = 0.
$$
Applying the Laplace expansion to the left-hand side, we obtain the
algebraic bilinear identity for determinants,
\begin{eqnarray}
\fl &|n-1, n+1,\cdots,n+N-2,n+N|\times
     |n,n+1,\cdots,n+N-2,n+N-1| \cr
\fl  - &|n, n+1,\cdots,n+N-2,n+N|\times
     |n-1,n+1,\cdots,n+N-2,n+N-1| \cr
\fl  + &|n+1,\cdots,n+N-2,n+N-1,n+N|\times
     |n-1,n,n+1,\cdots,n+N-2|
= 0,&
\end{eqnarray}
which is rewritten by using
(\ref{n-tau}), (\ref{x-dif}), (\ref{t-dif}) and (\ref{xt-dif}), into
the differential bilinear equation,
$$
(\partial_t(\partial_x-Nc)-1)\tau_n\times\tau_n
-(\partial_x-Nc)\tau_n\times\partial_t\tau_n+\tau_{n+1}\tau_{n-1}= 0\,,
$$
i.e.,
$$
\left(\partial_t \partial_x \tau_n-\tau_n
\right) \tau_n
-\partial_t \tau_n \partial_x \tau_n
 +\tau_{n+1}\tau_{n-1}
= 0\,.
$$
Setting $n=0$, $f=\tau_0\,,g=\tau_1\,,
h=\tau_{-1}\,$,
the above bilinear equation leads to the first equation of
(\ref{2dbilinear}). \\
\quad \\
{\bf The second equation of eqs.~(\ref{2dbilinear})}\\
Let us introduce two identities for $2N\times 2N$ determinants,
$$
\left|\matrix{
n-1&\vbl4  &n+1 & \cdots &n+N-3 &n+N-1 &\vbl4 &n+N &\vbl4 & n &\vbl4
& &\hbox{\O} & &\cr
\multispan{15}\hblfil \cr
n-1&\vbl4  & & &\hbox{\O} & &\vbl4 &n+N &\vbl4
& n &\vbl4  &n+1 &\cdots &n+N-1\cr}
 \right| = 0\,,
$$
$$
\left|\matrix{
n-1&\vbl4  &n+1 & \cdots &n+N-2 &\vbl4 &n+N+1
&\vbl4 &  & &\hbox{\O} & &\vbl4 &n+N-1 \cr
\multispan{14}\hblfil \cr
n-1&\vbl4  & &\hbox{\O} & &\vbl4 &n+N+1 &\vbl4
& n &n+1 &\cdots &n+N-2 &\vbl4 &n+N-1\cr}
 \right| = 0\,.
$$
Applying the Laplace expansion to the left-hand side, we obtain the
algebraic bilinear identities for determinants,
\begin{eqnarray}
&& \fl \quad |n-1, n+1,\cdots,n+N-3,n+N-1,n+N|\times
     |n,n+1,\cdots,n+N-2,n+N-1| \nonumber\\
&& \fl  - |n,n+1,\cdots,n+N-3,n+N-1,n+N|\times
     |n-1,n+1,\cdots,n+N-2,n+N-1| \nonumber\\
&& \fl  + |n-1,n,n+1,\cdots,n+N-3,n+N-1|\times
     |n+1,n+2,\cdots,n+N-1,n+N|
= 0,\nonumber\\
\end{eqnarray}
\begin{eqnarray}
&& \fl \quad |n-1,n+1,\cdots,n+N-2,n+N+1|\times
     |n,n+1,\cdots,n+N-2,n+N-1| \nonumber\\
&& \fl  - |n,n+1,\cdots,n+N-2,n+N+1|\times
     |n-1,n+1,\cdots,n+N-2,n+N-1| \nonumber\\
&& \fl  + |n+1,\cdots,n+N-1,n+N+1|\times
     |n-1,n,n+1,\cdots,n+N-3,n+N-2|
= 0\,.\nonumber\\
\end{eqnarray}
Taking the difference of these two bilinear identities,
it is rewritten by using
(\ref{n-tau})-(\ref{t-dif}) and (\ref{yt-dif}) into
the differential bilinear equation,
\begin{eqnarray*}
&& \partial_t(\partial_y-2c\partial_x+Nc^2)\tau_n\times\tau_n
-(\partial_y-2c\partial_x+Nc^2)\tau_n\times\partial_t\tau_n \\
&& \qquad +(\partial_x-Nc)\tau_{n+1}\times\tau_{n-1}
-\tau_{n+1}(\partial_x-Nc)\tau_{n-1}=0\,,
\end{eqnarray*}
i.e.,
$$
(\partial_t \partial_y \tau_n) \tau_n-\partial_y \tau_n \partial_t\tau_n
-2c((\partial_t \partial_x \tau_n)\tau_n -\partial_x \tau_n\partial_t \tau_n)
+(\partial_x\tau_{n+1})\tau_{n-1}-\tau_{n+1}(\partial_x\tau_{n-1})= 0\,,
$$
which is nothing but the second equation of (\ref{2dbilinear}).\\
\quad \\
{\bf The third equation of eqs.~(\ref{2dbilinear})}\\
Let us introduce two identities for $2N\times 2N$ determinants,
$$
\left|\matrix{
n-2 &n+1 & \cdots &n+N-2 &\vbl4 &n+N &\vbl4 &n
&\vbl4 & &\hbox{\O} & &\vbl4 &n+N-1 \cr
\multispan{14}\hblfil \cr
& &\hbox{\O} & &\vbl4 &n+N &\vbl4 & n &\vbl4
&n+1 &\cdots &n+N-2 &\vbl4 &n+N-1\cr}
 \right| = 0,
$$
$$
\left|\matrix{
n-1 &n& n+2 & \cdots &n+N-2 &\vbl4 &n+N &\vbl4 &n+1
&\vbl4
& & &\hbox{\O} & &\vbl4 &n+N-1 \cr
\multispan{16}\hblfil \cr
& & &\hbox{\O}    &    &\vbl4 &n+N &\vbl4 & n+1 &\vbl4 &n
&n+2 &\cdots &n+N-2 &\vbl4 &n+N-1\cr}
 \right| = 0.
$$
Applying the Laplace expansion to the left-hand side, we obtain the
algebraic bilinear identities for determinants,
\begin{eqnarray}
&& \fl \quad |n-2, n+1,n+2,\cdots,n+N-2,n+N|\times
     |n,n+1,\cdots,n+N-2,n+N-1| \nonumber\\
&& \fl  - |n-2,n+1,n+2,\cdots,n+N-2,n+N-1|\times
     |n,n+1,n+2,\cdots,n+N-2,n+N| \nonumber\\
&& \fl  + |n-2,n,n+1,n+2,\cdots,n+N-2|\times
     |n+1,n+2,\cdots,n+N-2,n+N-1,n+N|
= 0,\nonumber\\
\end{eqnarray}
\begin{eqnarray}
&& \fl \quad |n-1, n,n+2,\cdots,n+N-2,n+N|\times
     |n,n+1,\cdots,n+N-2,n+N-1| \nonumber\\
&& \fl  - |n-1,n,n+2,\cdots,n+N-2,n+N-1|\times
     |n,n+1,n+2,\cdots,n+N-2,n+N| \nonumber\\
&& \fl  + |n-1,n,n+1,n+2,\cdots,n+N-2|\times
     |n,n+2,\cdots,n+N-2,n+N-1,n+N|
= 0\,.\nonumber\\
\end{eqnarray}
Taking the difference of these two bilinear identities,
it is rewritten by using
(\ref{s-dif}) and (\ref{xs-dif}) into
the differential bilinear equation,
$$
\partial_s(\partial_x-Nc)\tau_n\times\tau_n
-\partial_s\tau_n\times(\partial_x-Nc)\tau_n
+(\partial_t\tau_{n-1})\tau_{n+1}-\tau_{n-1}(\partial_t\tau_{n+1})= 0\,,
$$
i.e.,
$$
(\partial_s \partial_x \tau_n) \tau_n
-\partial_s \tau_n \partial_x \tau_n
-(\partial_t\tau_{n+1})\tau_{n-1}+\tau_{n+1}(\partial_t\tau_{n-1})= 0\,.
$$
Setting $n=0$, $f=\tau_0\,,g=\tau_1\,,
h=\tau_{-1}\,$,
the above bilinear equation leads to the third equation in
(\ref{2dbilinear}). \\
\quad \\
{\bf The fourth equation of eqs.~(\ref{2dbilinear})}\\
Let us introduce four identities for $2N\times 2N$ determinants,
$$
\left|\matrix{
n-2 &\vbl4 &n+1 &\cdots &n+N-2 &\vbl4 &n+N+1 &\vbl4
 &n &\vbl4 &    &       &\hbox{\O} & \cr
\multispan{14}\hblfil \cr
n-2 &\vbl4 & &\hbox{\O} &      &\vbl4 &n+N+1 &\vbl4
 &n &\vbl4 &n+1 &\cdots &n+N-2 &n+N-1\cr}
 \right| = 0,
$$
$$
\left|\matrix{
n-2 &\vbl4 &n+1 &\cdots &n+N-3 &n+N-1 &\vbl4 &n+N &\vbl4
 &n &\vbl4 &    &\hbox{\O} & \cr
\multispan{14}\hblfil \cr
n-2 &\vbl4 &    &       &\hbox{\O} &  &\vbl4 &n+N &\vbl4
 &n &\vbl4 &n+1 &\cdots &n+N-1\cr}
 \right| = 0,
$$
$$
\left|\matrix{
n-1 &\vbl4 &n &n+2 &\cdots &n+N-2 &\vbl4 &n+N+1 &\vbl4
 &n+1 &\vbl4 &  &   &\hbox{\O} &  \cr
\multispan{15}\hblfil \cr
n-1 &\vbl4 &  & &\hbox{\O} &      &\vbl4 &n+N+1 &\vbl4
 &n+1 &\vbl4 &n &n+2 &\cdots &n+N-1\cr}
 \right| = 0,
$$
$$
\left|\matrix{
n-1 &\vbl4 &n &n+2 &\cdots &n+N-3 &n+N-1 &\vbl4 &n+N &\vbl4
 &n+1 &\vbl4 &  &   &\hbox{\O} &  \cr
\multispan{16}\hblfil \cr
n-1 &\vbl4 &  & &\hbox{\O} &      &      &\vbl4 &n+N &\vbl4
 &n+1 &\vbl4 &n &n+2 &\cdots &n+N-1\cr}
 \right| = 0.
$$
Applying the Laplace expansion to the left-hand side, we obtain the
algebraic bilinear identities for determinants,
\begin{eqnarray}
&& \fl \quad |n-2,n+1,\cdots,n+N-2,n+N+1|\times
     |n,n+1,\cdots,n+N-2,n+N-1| \nonumber\\
&& \fl  - |n,n+1,\cdots,n+N-2,n+N+1|\times
     |n-2,n+1,\cdots,n+N-2,n+N-1| \nonumber\\
&& \fl  + |n-2,n,n+1,\cdots,n+N-2|\times
     |n+1,\cdots,n+N-2,n+N-1,n+N+1|
= 0,\nonumber\\
\end{eqnarray}
\begin{eqnarray}
&& \fl \quad |n-2,n+1,\cdots,n+N-3,n+N-1,n+N|\times
     |n,n+1,\cdots,n+N-1| \nonumber\\
&& \fl  - |n,n+1,\cdots,n+N-3,n+N-1,n+N|\times
     |n-2,n+1,\cdots,n+N-1| \nonumber\\
&& \fl  + |n-2,n,n+1,\cdots,n+N-3,n+N-1|\times
     |n+1,\cdots,n+N-1,n+N|
= 0,\nonumber\\
\end{eqnarray}
\begin{eqnarray}
&& \fl \quad |n-1,n,n+2,\cdots,n+N-2,n+N+1|\times
     |n,n+1,\cdots,n+N-1| \nonumber\\
&& \fl  - |n,n+1,\cdots,n+N-2,n+N+1|\times
     |n-1,n,n+2,\cdots,n+N-1| \nonumber\\
&& \fl  + |n-1,n,n+1,\cdots,n+N-2|\times
     |n,n+2,\cdots,n+N-1,n+N+1|
= 0,\nonumber\\
\end{eqnarray}
\begin{eqnarray}
&& \fl \quad |n-1,n,n+2,\cdots,n+N-3,n+N-1,n+N|\times
     |n,n+1,\cdots,n+N-1| \nonumber\\
&& \fl  - |n,n+1,\cdots,n+N-3,n+N-1,n+N|\times
     |n-1,n,n+2,\cdots,n+N-1| \nonumber\\
&& \fl  + |n-1,n,n+1,\cdots,n+N-3,n+N-1|\times
     |n,n+2,\cdots,n+N-1,n+N|
= 0.\nonumber\\
\end{eqnarray}
Taking an appropriate linear combination of these four bilinear identities,
it is rewritten by using (\ref{ys-dif}) into
the differential bilinear equation,
\begin{eqnarray*}
&&\fl (\partial_s(\partial_y-2c\partial_x+Nc^2)-2)\tau_n\times\tau_n
-(\partial_y-2c\partial_x+Nc^2)\tau_n\times\partial_s\tau_n
+\partial_t\tau_{n-1}\times(\partial_x-Nc)\tau_{n+1} \\
&&\fl -(\partial_t(\partial_x-Nc)-1)\tau_{n-1}\times\tau_{n+1}
-\tau_{n-1}(\partial_t(\partial_x-Nc)-1)\tau_{n+1}
+(\partial_x-Nc)\tau_{n-1}\times\partial_t\tau_{n+1}= 0\,,
\end{eqnarray*}
i.e.,
\begin{eqnarray*}
&&\fl (\partial_s\partial_y\tau_n)\tau_n
-\partial_y\tau_n\partial_s\tau_n
-2c((\partial_s\partial_x\tau_n)\tau_n-\partial_x\tau_n\partial_s\tau_n)
-2\tau_n\tau_n \\
&&\fl -(\partial_t\partial_x\tau_{n+1})\tau_{n-1}
+\partial_x\tau_{n+1}\partial_t\tau_{n-1}
+\partial_t\tau_{n+1}\partial_x\tau_{n-1}
-\tau_{n+1}\partial_t\partial_x\tau_{n-1}
+2\tau_{n+1}\tau_{n-1}
= 0,
\end{eqnarray*}
which leads to the fourth equation in (\ref{2dbilinear}).
\end{proof}
\begin{theorem}
Bilinear equations
\begin{equation}
\left\{\begin{array}{l}\displaystyle
-\left(\frac{1}{2}D_tD_x-1\right)f\cdot f=gh\,,
\\[5pt]
2cff=(D_x+2c)g\cdot h\,,
\\[5pt]
-2ff=(D_tD_x+2cD_t-2)g\cdot h\,,
\end{array}\right.
\end{equation}
have a determinant solution
$$
f=\tau_0\,,
\qquad
g=\tau_1\,,
\qquad
h=\tau_{-1}\,,
$$
$$
\tau_n=\left|\matrix{
\psi_1^{(n)} &\psi_1^{(n+1)} &\cdots &\psi_1^{(n+N-1)} \cr
\psi_2^{(n)} &\psi_2^{(n+1)} &\cdots &\psi_2^{(n+N-1)} \cr
\vdots       &\vdots         &       &\vdots           \cr
\psi_N^{(n)} &\psi_N^{(n+1)} &\cdots &\psi_N^{(n+N-1)}}\right|\,,
$$
where
$$
\psi_i^{(n)}=a_{i,1}(p_i-c)^ne^{\xi_i}+a_{i,2}(-p_i-c)^ne^{\eta_i}\,,
$$
$$
\xi_i=p_ix+\frac{1}{p_i-c}t+\frac{1}{(p_i-c)^2}s+\xi_{i0}\,,
$$
$$
\eta_i=-p_ix-\frac{1}{p_i+c}t+\frac{1}{(p_i+c)^2}s+\eta_{i0}\,.
$$
\end{theorem}
\begin{proof}
In the previous Lemma, apply the 2-reduction condition
$$
q_i=-p_i\,.
$$
This condition gives a constraint
$$
D_y=0\,,
$$
into bilinear equations.
Thus we have
\begin{equation}
\left\{\begin{array}{l}\displaystyle
\left(\frac{1}{2}D_tD_x-1\right)f\cdot f=-gh\,,
\\[5pt]
cD_tD_xf\cdot f=D_xg\cdot h\,,
\\[5pt]\displaystyle
\frac{1}{2}D_sD_xf\cdot f=D_tg\cdot h\,,
\\[5pt]
(-cD_sD_x-2)f\cdot f=(D_tD_x-2)g\cdot h\,.
\end{array}\right.
\end{equation}
After simple manipulations, we have
\begin{equation}
\left\{\begin{array}{l}\displaystyle
-\left(\frac{1}{2}D_tD_x-1\right)f\cdot f=gh\,,
\\[5pt]
2cff=(D_x+2c)g\cdot h\,,
\\[5pt]
-2ff=(D_tD_x+2cD_t-2)g\cdot h\,.
\end{array}\right.
\end{equation}
Let us consider a determinant solution.
When we apply the 2-reduction condition
$$
q_i=-p_i\,,
$$
on the determinant solution in the previous Lemma,
we will have
$$
\tau_n=\left|\matrix{
\psi_1^{(n)} &\psi_1^{(n+1)} &\cdots &\psi_1^{(n+N-1)} \cr
\psi_2^{(n)} &\psi_2^{(n+1)} &\cdots &\psi_2^{(n+N-1)} \cr
\vdots       &\vdots         &       &\vdots           \cr
\psi_N^{(n)} &\psi_N^{(n+1)} &\cdots &\psi_N^{(n+N-1)}}\right|\,,
$$
where
$$
\psi_i^{(n)}=a_{i,1}(p_i-c)^ne^{\xi_i}+a_{i,2}(-p_i-c)^ne^{\eta_i}\,,
$$
$$
\xi_i=p_ix+\frac{1}{p_i-c}t+\frac{1}{(p_i-c)^2}s+\xi_{i0}\,,
$$
$$
\eta_i=-p_ix-\frac{1}{p_i+c}t+\frac{1}{(p_i+c)^2}s+\eta_{i0}\,.
$$
Thus the theorem was proved.
\end{proof}

\begin{theorem}
\label{ch:determinant}
The CH equation
$$
(\partial_T+w\partial_X)(w_{XX}-w)
+2w_X\left(w_{XX}-w-\kappa^2 \right)=0\,,
$$
i.e.,
$$
w_T+2\kappa^2w_X-w_{TXX}+3ww_X=2w_Xw_{XX}+ww_{XXX},
$$
where $\kappa^2=1/c$,
is decomposed into bilinear equations
\begin{equation}
\left\{\begin{array}{l}\displaystyle
-\left(\frac{1}{2}D_tD_x-1\right)f\cdot f=gh\,,
\\[5pt]
2cff=(D_x+2c)g\cdot h\,,
\\[5pt]
-2ff=(D_tD_x+2cD_t-2)g\cdot h\,,
\end{array}\right.\label{ch-bilinear}
\end{equation}
through the hodograph transformation
$$
\left\{\begin{array}{l}
X=2cx+\log \displaystyle{\frac{g}{h}}\,,
\\
T=t\,,
\end{array}\right.
$$
and the dependent variable transformation
$$
w=\left(\log\frac{g}{h}\right)_t\,.
$$
\end{theorem}
\begin{proof}
Consider the dependent variable transformation
$$
u=\frac{g}{f}\,,
\qquad
v=\frac{h}{f}\,.
$$
{}From bilinear equations (\ref{ch-bilinear}), we obtain
\begin{equation}
\left\{\begin{array}{l}
-((\log f)_{xt}-1)=uv\,,
\\[5pt]
2c=(D_x+2c)u\cdot v\,,
\\[5pt]
-2=(D_tD_x+(2\log f)_{xt}+2cD_t-2)u\cdot v\,.
\end{array}\right.\label{chbi1}
\end{equation}
Eliminating $(\log f)_{xt}$ from the third equation
using the first equation in
(\ref{chbi1}), we obtain
\begin{equation}
\left\{\begin{array}{l}
2c=(D_x+2c)u\cdot v\,,
\\[5pt]
-2=(D_tD_x+2cD_t-2uv)u\cdot v\,.
\end{array}\right.
\end{equation}
Thus we have
\begin{equation}
\left\{\begin{array}{l}\displaystyle
2c=\left(\left(\log\frac{u}{v}\right)_x+2c\right)uv\,,
\\[5pt]\displaystyle
-2=\left((\log uv)_{xt}+\left(\log\frac{u}{v}\right)_x
\left(\log\frac{u}{v}\right)_t
+2c\left(\log\frac{u}{v}\right)_t-2uv\right)uv\,.
\end{array}\right.\label{chbi2}
\end{equation}
Eliminating $\left(\log\frac{u}{v}\right)_x
+2c$ from the
second equation using the first equation in (\ref{chbi2}), we have
$$
\left\{\begin{array}{l}\displaystyle
2c=\left(\left(\log\frac{u}{v}\right)_x+2c\right)uv\,,
\\[5pt]\displaystyle
-2=\left((\log uv)_{xt}+\frac{2c}{uv}\left(\log\frac{u}{v}\right)_t
-2uv\right)uv\,.
\end{array}\right.
$$
Introducing new dependent variables
$$
\phi=\frac{u}{v}=\frac{g}{h}\,,
\qquad
\rho=uv=\frac{gh}{f^2}\,,
$$
we have
\begin{equation}
\left\{\begin{array}{l}\displaystyle
\frac{2c}{\rho}=(\log\phi)_x+2c\,,
\\[5pt]
-2=\rho(\log\rho)_{xt}+2c(\log\phi)_t-2\rho^2\,.
\end{array}\right.\label{phirho}
\end{equation}
%%%\begin{equation}
%%%\left\{\begin{array}{l}\displaystyle
%%%\frac{2c}{\rho}=(\log\phi)_x+2c
%%%\\[5pt]\displaystyle
%%%\frac{\rho}{2c}(\log\rho)_{xt}+(\log\phi)_t+\frac{1}{c}=\frac{1}{c}\rho^2
%%%\end{array}\right.\label{phirho}
%%%\end{equation}
Let
$$
w=(\log\phi)_t=\left(\log\frac{g}{h}\right)_t\,.
$$
{}From the first equation in (\ref{phirho}), we obtain
$$
-(\log\rho)_t=\frac{(\log\phi)_{xt}}{2c+(\log\phi)_x}
=\frac{\rho}{2c}(\log\phi)_{xt}=\frac{\rho}{2c}w_x\,.
$$
Thus we have
\begin{equation}
\left\{\begin{array}{l}\displaystyle
-(\log\rho)_t=\frac{\rho}{2c}w_x\,,
\\[5pt]\displaystyle
\frac{\rho}{2c}(\log\rho)_{xt}+w+\frac{1}{c}=\frac{1}{c}\rho^2\,.
\end{array}\right.
\end{equation}
Replacing $(\log\rho)_{t}$ in the second equation
by $-\frac{\rho}{2c}w_x$ using the first equation, we obtain
$$
\left\{\begin{array}{l}\displaystyle
-(\log\rho)_t=\frac{\rho}{2c}w_x\,,
\\[5pt]\displaystyle
-\frac{\rho}{2c}\left(\frac{\rho}{2c}w_x\right)_x
+w+\frac{1}{c}=\frac{1}{c}\rho^2\,.
\end{array}\right.
$$
Consider the hodograph transformation
$$
\left\{\begin{array}{l}
X=2cx+\log\phi\,,
\\
T=t\,.
\end{array}\right.
$$
Then we have
$$
\frac{\partial X}{\partial x}=2c+\left(\log\phi\right)_x
=2c+\left(\log\frac{g}{h}\right)_x
=2c\frac{f^2}{gh}=\frac{2c}{\rho}\,,
\qquad
\frac{\partial X}{\partial t}=(\log\phi)_t=\left(\log\frac{g}{h}\right)_t=w\,,
$$
$$
\left\{\begin{array}{l}\displaystyle
\partial_x=\frac{2c}{\rho}\partial_X\,,
\\[5pt]
\partial_t=\partial_T+w\partial_X\,.
\end{array}\right.
$$
Using these results, we obtain
\begin{equation}
\left\{\begin{array}{l}\displaystyle
-(\partial_T+w\partial_X)\log\rho=w_X\,,
\\[5pt]\displaystyle
-w_{XX}+w+\frac{1}{c}=\frac{1}{c}\rho^2\,.
\end{array}\right.\label{ch-derive}
\end{equation}
Eliminating $\rho$ from the first equation using the second equation in
eqs.(\ref{ch-derive}), we obtain
$$
(\partial_T+w\partial_X)\log\left(-w_{XX}+w+\frac{1}{c}\right)=-2w_X\,.
$$
Thus we finally obtain the CH equation
$$
(\partial_T+w\partial_X)(w_{XX}-w)+2w_X\left(w_{XX}-w-\frac{1}{c}\right)=0\,.
$$
The theorem was approved.
\end{proof}
Furthermore we have the following theorem.
\begin{corollary}
The CH equation has a determinant form of $N$-soliton solutions.
\end{corollary}
\begin{proof}
{}From Theorem 2.2 and 2.3, the proof is obvious.
\end{proof}

\section{A semi-discrete Camassa-Holm equation}
\noindent

\begin{lemma}
Bilinear equations
\begin{equation}
\left\{\begin{array}{l}\displaystyle
\left(\frac{1-ac}{a}D_t-1\right)f(k+1,l)\cdot f(k,l)=-g(k+1,l)h(k,l)\,,
\\[5pt]\displaystyle
\left(\frac{1-bc}{b}D_t-1\right)f(k,l+1)\cdot f(k,l)=-g(k,l+1)h(k,l)\,,
\\[5pt]\displaystyle
\left(\frac{1-ac}{a}D_s-D_t\right)f(k+1,l)\cdot f(k,l)=D_tg(k+1,l)\cdot h(k,l)\,,
\\[5pt]\displaystyle
\left(\frac{1-bc}{b}D_s-D_t\right)f(k,l+1)\cdot f(k,l)=D_tg(k,l+1)\cdot h(k,l)\,,
\end{array}\right.\label{semi-bilinear}
\end{equation}
have a determinant solution
$$
f(k,l)=\tau_0(k,l)\,,
\qquad
g(k,l)=\tau_1(k,l)\,,
\qquad
h(k,l)=\tau_{-1}(k,l)\,,
$$
$$
\tau_n(k,l)=\left|\matrix{
\psi_1^{(n)}(k,l) &\psi_1^{(n+1)}(k,l) &\cdots &\psi_1^{(n+N-1)}(k,l) \cr
\psi_2^{(n)}(k,l) &\psi_2^{(n+1)}(k,l) &\cdots &\psi_2^{(n+N-1)}(k,l) \cr
\vdots            &\vdots              &       &\vdots                \cr
\psi_N^{(n)}(k,l) &\psi_N^{(n+1)}(k,l) &\cdots &\psi_N^{(n+N-1)}(k,l)}\right|\,,
$$
where
$$
\psi_i^{(n)}(k,l)=a_{i,1}(p_i-c)^n(1-ap_i)^{-k}(1-bp_i)^{-l}e^{\xi_i}
+a_{i,2}(q_i-c)^n(1-aq_i)^{-k}(1-bq_i)^{-l}e^{\eta_i}\,,
$$
$$
\xi_i=\frac{1}{p_i-c}t+\frac{1}{(p_i-c)^2}s+\xi_{i0}\,,
$$
$$
\eta_i=\frac{1}{q_i-c}t+\frac{1}{(q_i-c)^2}s+\eta_{i0}\,.
$$
\end{lemma}
\begin{proof}
Consider the following Casorati determinant solution,
$$
\tau_n(k,l)=\left|\matrix{
\psi_1^{(n)}(k,l) &\psi_1^{(n+1)}(k,l) &\cdots &\psi_1^{(n+N-1)}(k,l) \cr
\psi_2^{(n)}(k,l) &\psi_2^{(n+1)}(k,l) &\cdots &\psi_2^{(n+N-1)}(k,l) \cr
\vdots            &\vdots              &       &\vdots                \cr
\psi_N^{(n)}(k,l) &\psi_N^{(n+1)}(k,l) &\cdots
 &\psi_N^{(n+N-1)}(k,l)}\right|
\,,
$$
where $\psi_i^{(n)}$'s are arbitrary functions of two continuous
independent variables, $t$ and $s$, and two discrete independent
 variables, $k$ and $l$, which
satisfy the linear dispersion relations,
\begin{eqnarray}
&&\Delta_k\psi_i^{(n)} = \psi_i^{(n+1)}+c\psi_i^{(n)}\,,
\label{k-dispersion-d}\\
&&\Delta_l \psi_i^{(n)} =\psi_i^{(n+1)}+c\psi_i^{(n)} \,,
\label{l-dispersion-d}\\
&&
\partial_t\psi_i^{(n)} = \psi_i^{(n-1)}\,,
\label{t-linear-d}\\
&&
\partial_s\psi_i^{(n)} = \psi_i^{(n-2)}\,,
\label{s-linear-d}
\end{eqnarray}
where $\Delta_k$ and $\Delta_l$ are
defined as $\Delta_k \psi(k,l)=\frac{\psi(k,l)-\psi(k-1,l)}{a}$ and 
$\Delta_l \psi(k,l)=\frac{\psi(k,l)-\psi(k,l-1)}{b}$, respectively.
Thus we can choose $\psi_i^{(n)}$ as follows:
$$
\psi_i^{(n)}(k,l)=(p_i-c)^n(1-ap_i)^{-k}(1-bp_i)^{-l}e^{\xi_i}
+(q_i-c)^n(1-aq_i)^{-k}(1-bq_i)^{-l}e^{\eta_i}\,,
$$
$$
\xi_i=\frac{1}{p_i-c}t+\frac{1}{(p_i-c)^2}s+\xi_{i0}\,,
$$
$$
\eta_i=\frac{1}{q_i-c}t+\frac{1}{(q_i-c)^2}s+\eta_{i0}\,.
$$

We use the following notation for
simplicity:
\begin{eqnarray*}
|{n_{1_{k_1,l_1}}},
  {n_{2_{k_2,l_2}}},
  \cdots,
  {n_{N_{k_N,l_N}}}|
 = \left|\matrix{
  \psi_1^{(n_1)}(k_1,l_1) &\psi_1^{(n_2)}(k_2,l_2)&\cdots
   &\psi_1^{(n_N)}(k_N,l_N)\cr
  \psi_2^{(n_1)}(k_1,l_1) &\psi_2^{(n_2)}(k_2,l_2)&\cdots
   &\psi_2^{(n_N)}(k_N,l_N) \cr
  \vdots                     &\vdots                     &
   &\vdots                     \cr
  \psi_N^{(n_1)}(k_1,l_1) &\psi_N^{(n_2)}(k_2,l_2) &\cdots
   &\psi_N^{(n_N)}(k_N,l_N) \cr}
 \right|.
\end{eqnarray*}
In this notation, $\tau_n(k,l)$ is rewritten as 
\[
\tau_n(k,l)=|n_{k,l},n+1_{k,l},\cdots ,n+N-1_{k,l}|\,,
\]
or suppressing the index $k$ and $l$,
\[
\tau_n(k,l)=|n,n+1,\cdots ,n+N-1|\,. 
\]
\quad \\
{\bf The first equation of eqs.~(\ref{semi-bilinear})}\\
By using the Casoratian technique developed in \cite{OHTI} and \cite{OKMS},
it is possible to derive the following differential and difference
formulas for the $\tau$ function,
\begin{eqnarray}
&&\tau_{n-1}(k,l)=|n-1,n,\cdots ,n+N-2|,\\
&&\tau_n(k+1,l)
=\frac{1}{(1-ac)^{N-2}}
|n_{k+1},n+1,\cdots ,n+N-2,n+N-1_{k+1}|,\nonumber \\
&&\quad =\frac{1}{(1-ac)^{N-1}}
|n,\cdots ,n+N-2,n+N-1_{k+1}|,\\
&&\tau_{n+1}(k+1,l)
=\frac{1}{(1-ac)^{N-1}}|n+1,\cdots
 ,n+N-1,n+N_{k+1}|,\nonumber\\
&&\quad =\frac{1}{a(1-ac)^{N-2}}|n+1,\cdots
 ,n+N-1,n+N-1_{k+1}|,\\
&&\partial_t \tau_n =
|n-1,
n+1,
\cdots,n+N-2,
n+N-1|\,,\\
&&\frac{1-ac}{a}\tau_n(k+1,l)
=\frac{1}{a(1-ac)^{N-2}}|n,\cdots ,n+N-2,n+N-1_{k+1}|\,,\\
&&\frac{1-ac}{a}\partial_t\tau_n(k+1,l)-\tau_n(k+1,l)\nonumber\\
&&\quad =\frac{1-ac}{a(1-ac)^{N-2}} |n-1_{k+1},n+1,
\cdots,n+N-2,n+N-1_{k+1}|\nonumber\\
&&\quad -|n_{k+1},n+1_{k+1},
\cdots,n+N-2_{k+1},n+N-1_{k+1}|\,\nonumber\\
&&\quad =
\frac{1}{a(1-ac)^{N-2}} |n-1,n+1,
\cdots,n+N-2,n+N-1_{k+1}|\nonumber\\
&&\quad +\frac{1}{(1-ac)^{N-2}} |n_{k+1},n+1,
\cdots,n+N-2,n+N-1_{k+1}|\nonumber\\
&&\quad -\frac{1}{(1-ac)^{N-2}}|n_{k+1},n+1,
\cdots,n+N-2,n+N-1_{k+1}|\nonumber\\
&&\quad =\frac{1}{a(1-ac)^{N-2}} |n-1,n+1,
\cdots,n+N-2,n+N-1_{k+1}|\,,
\end{eqnarray}
Let us introduce an identity for $2N\times 2N$ determinant,
$$
\left|\matrix{
n+1_{k} & \cdots &n+N-2_{k} &\vbl4 &n+N-1_{k} &\vbl4 &n+N-1_{k+1}
&\vbl4
& n-1_{k} &\vbl4 & n_{k} &\vbl4 & & \hbox{\O} & \cr
\multispan{15}\hblfil \cr
&\hbox{\O}    &    &\vbl4 &n+N-1_{k} &\vbl4 & n+N-1_{k+1} &\vbl4
& n-1_{k} &\vbl4 & n_{k} &\vbl4 & n+1_{k}  &\cdots &n+N-2_{k} \cr}
 \right| = 0.
$$
Applying the Laplace expansion to the left-hand side, we obtain the
algebraic bilinear identity for determinants,
\begin{eqnarray}
\fl &|n-1, n+1,\cdots,n+N-2,n+N-1_{k+1}|\times
     |n,n+1,\cdots,n+N-2,n+N-1| \cr
\fl  - &|n, n+1,\cdots,n+N-2,n+N-1_{k+1}|\times
     |n-1,n+1,\cdots,n+N-2,n+N-1| \cr
\fl  + &|n+1,\cdots,n+N-2,n+N-1,n+N-1_{k+1}|\times
     |n-1,n,n+1,\cdots,n+N-2|
= 0,&
\end{eqnarray}
which is rewritten into
the differential bilinear equation,
$$
\left(\frac{1-ac}{a}\partial_t \tau_n(k+1,l)
-\tau_n(k+1,l)\right) \tau_n(k,l)
-\frac{1-ac}{a}\tau_n(k+1,l)\partial_t \tau_n(k,l) +
\tau_{n+1}(k+1,l)\tau_{n-1}(k,l)= 0\,.
$$
Setting $n=0$, $f=\tau_0\,,g=\tau_1\,,
h=\tau_{-1}\,$,
the above bilinear equation leads to the first equation in
(\ref{semi-bilinear}). \\
\quad \\
{\bf The second equation of eqs.~(\ref{semi-bilinear})}\\
The proof is similar to the proof of the first equation.\\
\quad \\
{\bf The third equation of eqs.~(\ref{semi-bilinear})}\\
We use the following differential and difference
formulas for the $\tau$ function
%%%(use $\psi_i^{(n+1)}(k+1)-\frac{a}{1-ac}\psi_i^{(n)}(k+1)
%%%=\frac{1}{1-ac}\psi_i^{(n)}(k)$),
\begin{eqnarray}
&&\tau_{n-1}(k,l)=|n-1,n,\cdots ,n+N-2|,\\
&&\tau_n(k+1,l)
=
|n_{k+1},n+1_{k+1},\cdots ,n+N-2_{k+1},n+N-1_{k+1}|,\\
&&\tau_{n+1}(k+1,l)
=|n+1_{k+1},\cdots
 ,n+N-1_{k+1},n+N_{k+1}|,\nonumber\\
&&\quad =-\frac{1}{a}|n+1_{k+1},\cdots
 ,n+N-1_{k+1},n+N-1|,\\
&&\partial_t \tau_n(k,l) =
|n-1,
n+1,
\cdots,n+N-2,
n+N-1|\,,\\
&&\frac{1-ac}{a}\tau_n(k+1,l)
=\frac{1}{a}|n,n+1_{k+1},\cdots ,n+N-2_{k+1},n+N-1_{k+1}|\nonumber\\
&& \quad
=\frac{1}{a}|n_{k+1},n+1,n+2_{k+1},\cdots ,n+N-2_{k+1},n+N-1_{k+1}|\,,\\
&&\partial_t \tau_n(k+1,l) =
|n-1_{k+1},
n+1_{k+1},
\cdots,n+N-2_{k+1},
n+N-1_{k+1}|\,,\\
&&\partial_t \tau_{n+1}(k+1,l) =
|n_{k+1},
n+2_{k+1},
\cdots,n+N-1_{k+1},
n+N_{k+1}|\\
&&\quad =
-\frac{1}{a}
|n_{k+1},
n+2_{k+1},
\cdots,n+N-1_{k+1},
n+N-1|
\,,\\
&&\partial_t \tau_{n-1}(k,l) =
|n-2,
n,
\cdots,n+N-3,
n+N-2|\,,\\
&&\partial_s \tau_n(k,l) =
|n-2,
n+1,
\cdots,n+N-2,
n+N-1|\nonumber\\
&&\quad +|n,
n-1,n+2,
\cdots,n+N-2,
n+N-1|
\,,\\
&&\frac{1-ac}{a}\partial_s \tau_n(k+1,l) =
\frac{1-ac}{a}|n-2_{k+1},
n+1_{k+1},
\cdots,n+N-2_{k+1},
n+N-1_{k+1}|\quad \nonumber \\
&&\quad +\frac{1-ac}{a}|n_{k+1},
n-1_{k+1},n+2_{k+1},
\cdots,n+N-2_{k+1},
n+N-1_{k+1}|
\,,\nonumber\\
&&\quad =
|n-1_{k+1},
n+1_{k+1},
\cdots,n+N-2_{k+1},
n+N-1_{k+1}|\nonumber \\
&&\quad +
\frac{1}{a}|n-2,
n+1_{k+1},
\cdots,n+N-2_{k+1},
n+N-1_{k+1}|\nonumber \\
&&\quad -\frac{1}{a}|n-1, n_{k+1},
n+2_{k+1},
\cdots,n+N-2_{k+1},
n+N-1_{k+1}|
\,,
\end{eqnarray}

Let us introduce following identities for $2N\times 2N$ determinant,
$$\small
\left|\matrix{
n-1_{k}  &\vbl4 & n_{k+1} & n+2_{k+1}
& \cdots &n+N-1_{k+1} &\vbl4 & n_{k} &\vbl4
& n+1_{k}&\vbl4 & & \hbox{\O} &  &\vbl4 &n+N-1_{k} &\cr
\multispan{17}\hblfil \cr
n-1_{k} &\vbl4 &   &  & \hbox{\O}  &  &\vbl4 & n_{k} &\vbl4
& n+1_k & \vbl4 & n+2_{k}  &\cdots & n+N-2_k &\vbl4 & n+N-1_{k} &
\cr }
\right| = 0\,,
$$
and
$$
\left|\matrix{
n-2_{k} &\vbl4 & n+1_{k+1} &  \cdots &n+N-1_{k+1} &\vbl4
 & n_{k}
&\vbl4
& & \hbox{\O} &  &\vbl4 &n+N-1_{k} &\cr
\multispan{14}\hblfil \cr
n-2_{k} &\vbl4 &   & \hbox{\O}  &  &\vbl4 & n_{k}
&\vbl4
& n+1_k &\cdots & n+N-2_k &\vbl4 & n+N-1_{k} &
\cr }
\right| = 0\,.
$$
Applying the Laplace expansion to the left-hand side, we obtain the
algebraic bilinear identities for determinants,
we obtain the
algebraic bilinear identities for determinants,
\begin{eqnarray}
\fl &&|n-1, n_{k+1},n+2_{k+1}, \cdots,n+N-1_{k+1}|\times
     |n,n+1,n+2, \cdots,n+N-2,n+N-1| \cr
\fl  - &&|n, n_{k+1},n+2_{k+1},\cdots,n+N-1_{k+1}|\times
     |n-1,n+1,\cdots,n+N-2,n+N-1| \cr
\fl  + &&|n+1,n_{k+1},n+2_{k+1},\cdots,n+N-1_{k+1}|\times
     |n-1,n,n+2,\cdots,n+N-2,n+N-1|\cr
\fl  - &&|n_{k+1},n+2_{k+1},\cdots,n+N-1_{k+1},n+N-1|\times
     |n-1,n,n+1,n+2, \cdots,n+N-2|\cr
\fl &&= 0,\label{semiidentity-3-1}
\end{eqnarray}
and
\begin{eqnarray}
\fl &&|n-2, n+1_{k+1}, \cdots,n+N-1_{k+1}|\times
     |n,n+1,n+2, \cdots,n+N-2,n+N-1| \cr
\fl  - &&|n, n+1_{k+1},\cdots,n+N-1_{k+1}|\times
     |n-2,n+1,\cdots,n+N-2,n+N-1| \cr
\fl  + &&|n+1_{k+1},\cdots,n+N-1_{k+1},n+N-1|\times
     |n-2,n,n+1,\cdots,n+N-2|\cr
\fl &&= 0,\label{semiidentity-3-2}
\end{eqnarray}
%%%Computing (\ref{semiidentity-3-2}) $\times \frac{1}{a}
%%%-$ (\ref{semiidentity-3-1}) $\times \frac{1}{a}$,
%%%we will have a bilinear equation
By dividing the difference of above two equations by $a$,
we arrive at a bilinear equation
\begin{eqnarray*}
&&\frac{1-ac}{a}\partial_s \tau_n(k+1,l)\tau_n(k,l)
-\frac{1-ac}{a}\tau_n(k+1,l)\partial_s \tau_n(k,l)\\
&&-\partial_t \tau_n(k+1,l)\tau_n(k,l)
+\tau_n(k+1,l) \partial_t \tau_n(k,l)\\
&&-\partial_t \tau_{n+1}(k+1,l)\tau_{n-1}(k,l)
+\tau_{n+1}(k+1,l) \partial_t \tau_{n-1}(k,l)
= 0\,.
\end{eqnarray*}
Setting $n=0$, $f=\tau_0\,,g=\tau_1\,,
h=\tau_{-1}\,$,
the above bilinear equation leads to the third equation in
(\ref{semi-bilinear}). \\
\quad \\
{\bf The fourth equation of eqs.~(\ref{semi-bilinear})}\\
The proof is similar to the proof of the third equation.
\end{proof}
\begin{theorem}
Bilinear equations
\begin{equation}
\left\{\begin{array}{l}\displaystyle
\left(\frac{1-ac}{a}D_t-1\right)f_{k+1}\cdot f_k=-g_{k+1}h_k\,,
\\[5pt]\displaystyle
\left(\frac{1+ac}{a}D_t-1\right)f_{k+1}\cdot f_k=-g_kh_{k+1}\,,
\\[5pt]\displaystyle
\left(\frac{1-ac}{a}D_s-D_t\right)f_{k+1}\cdot f_{k}=D_tg_{k+1}\cdot h_k\,,
\\[5pt]\displaystyle
\left(\frac{1+ac}{a}D_s+D_t\right)f_{k+1}\cdot f_{k}=D_tg_{k}\cdot h_{k+1}\,,
\end{array}\right.
\end{equation}
have a determinant solution
$$
f(k,l)=\tau_0(k,0)\,,
\qquad
g(k,l)=\tau_1(k,0)\,,
\qquad
h(k,l)=\tau_{-1}(k,0)\,,
$$
$$
\tau_n(k,l)=\left|\matrix{
\psi_1^{(n)}(k,l) &\psi_1^{(n+1)}(k,l) &\cdots &\psi_1^{(n+N-1)}(k,l) \cr
\psi_2^{(n)}(k,l) &\psi_2^{(n+1)}(k,l) &\cdots &\psi_2^{(n+N-1)}(k,l) \cr
\vdots            &\vdots              &       &\vdots                \cr
\psi_N^{(n)}(k,l) &\psi_N^{(n+1)}(k,l) &\cdots &\psi_N^{(n+N-1)}(k,l)}\right|\,
$$
where
$$
\psi_i^{(n)}(k,l)=a_{i,1}(p_i-c)^n(1-ap_i)^{-k}(1+ap_i)^{-l}e^{\xi_i}
+a_{i,2}(-p_i-c)^n(1+ap_i)^{-k}(1-ap_i)^{-l}e^{\eta_i}\,,
$$
$$
\xi_i=\frac{1}{p_i-c}t+\frac{1}{(p_i-c)^2}s+\xi_{i0}\,,
$$
$$
\eta_i=-\frac{1}{p_i+c}t+\frac{1}{(p_i+c)^2}s+\eta_{i0}\,.
$$
\end{theorem}
\begin{proof}
Applying the 2-reduction condition
$$
q_i=-p_i\,,\quad b=-a\,,
$$
the $\tau$ function satisfies
$$
\tau_n(k+1,l+1)
=\frac{1}{\displaystyle\prod_{i=1}^N(1-ap_i)(1+ap_i)}\tau_n(k,l)\,.
$$
Let
$$
f_k=f(k,0)\,,
\qquad
g_k=g(k,0)\,,
\qquad
h_k=h(k,0)\,,
$$
Then
$$
\left\{\begin{array}{l}\displaystyle
\left(\frac{1-ac}{a}D_t-1\right)f_{k+1}\cdot f_k=-g_{k+1}h_k\,,
\\[5pt]\displaystyle
\left(\frac{1+ac}{a}D_t-1\right)f_{k+1}\cdot f_k=-g_kh_{k+1}\,,
\\[5pt]\displaystyle
\left(\frac{1-ac}{a}D_s-D_t\right)f_{k+1}\cdot f_{k}=D_tg_{k+1}\cdot h_k\,,
\\[5pt]\displaystyle
\left(\frac{1+ac}{a}D_s+D_t\right)f_{k+1}\cdot f_{k}=D_tg_{k}\cdot h_{k+1}\,.
\end{array}\right.
$$
Let us consider a determinant solution.
When we apply the 2-reduction condition
$$
q_i=-p_i\,,\quad  b=-a\,,
$$
on the determinant solution in the previous Lemma,
we will have
$$
\tau_n(k,l)=\left|\matrix{
\psi_1^{(n)}(k,l) &\psi_1^{(n+1)}(k,l) &\cdots &\psi_1^{(n+N-1)}(k,l) \cr
\psi_2^{(n)}(k,l) &\psi_2^{(n+1)}(k,l) &\cdots &\psi_2^{(n+N-1)}(k,l) \cr
\vdots            &\vdots              &       &\vdots                \cr
\psi_N^{(n)}(k,l) &\psi_N^{(n+1)}(k,l) &\cdots &\psi_N^{(n+N-1)}(k,l)}\right|\,
$$
where
$$
\psi_i^{(n)}(k,l)=a_{i,1}(p_i-c)^n(1-ap_i)^{-k}(1+ap_i)^{-l}e^{\xi_i}
+a_{i,2}(-p_i-c)^n(1+ap_i)^{-k}(1-ap_i)^{-l}e^{\eta_i}\,,
$$
$$
\xi_i=\frac{1}{p_i-c}t+\frac{1}{(p_i-c)^2}s+\xi_{i0}\,,
$$
$$
\eta_i=-\frac{1}{p_i+c}t+\frac{1}{(p_i+c)^2}s+\eta_{i0}\,.
$$
Thus the theorem was proved.
\end{proof}

We propose a semi-discrete analogue of the CH equation
$$
\left\{\begin{array}{l}\displaystyle
\Delta^2 w_k=\frac{1}{\delta_k}M\left(\delta_kMw_k+
\frac{\kappa^2}{\delta_k}
\frac{\displaystyle \kappa^4\delta_k^2-4a^2}{\kappa^4-a^2}\right)\,,
\\[5pt]\displaystyle
\partial_t\delta_k=\left(1-\frac{\delta_k^2}{4}\right)\delta_k\Delta w_k\,,
\end{array}\right.
$$
where
a difference operator
$\Delta$ and an average operator $M$ are defined as
$$
\Delta F_k=\frac{F_{k+1}-F_k}{\delta_k}\,,
\qquad
MF_k=\frac{F_{k+1}+F_k}{2}\,.
$$
\begin{theorem}
\label{dch:determinant}
The semi-discrete CH equation
\begin{equation}
\left\{\begin{array}{l}\displaystyle
\Delta^2 w_k=\frac{1}{\delta_k}M\left(\delta_kMw_k+
\frac{\kappa^2}{\delta_k}
\frac{\displaystyle \kappa^4\delta_k^2-4a^2}{\kappa^4-a^2}\right)\,,
\\[5pt]\displaystyle
\partial_t\delta_k=\left(1-\frac{\delta_k^2}{4}\right)\delta_k\Delta w_k\,,
\end{array}\right.\label{semi-d-ch}
\end{equation}
is decomposed into bilinear equations
\begin{equation}
\left\{\begin{array}{l}\displaystyle
\left(\frac{1-ac}{a}D_t-1\right)f_{k+1}\cdot f_k=-g_{k+1}h_k\,,
\\[5pt]\displaystyle
\left(\frac{1+ac}{a}D_t-1\right)f_{k+1}\cdot f_k=-g_kh_{k+1}\,,
\\[5pt]\displaystyle
\left(\frac{1-ac}{a}D_s-D_t\right)f_{k+1}\cdot f_{k}=D_tg_{k+1}\cdot h_k\,,
\\[5pt]\displaystyle
\left(\frac{1+ac}{a}D_s+D_t\right)f_{k+1}\cdot f_{k}=D_tg_{k}\cdot h_{k+1}\,,
\end{array}\right.
\label{bilinear-semi}
\end{equation}
through the transformation
$$
\delta_k=\frac{4a}{(\kappa^2+a)\frac{g_{k+1}h_k}{f_{k+1}f_k}
+(\kappa^2-a)\frac{g_{k}h_{k+1}}{f_{k+1}f_k}}\,,
$$
and
$$
w_k=\left(\log\frac{g_k}{h_k}\right)_t\,.
$$
where $\kappa^2=1/c$\,.
\end{theorem}
\begin{proof}
Let us start from bilinear equations (\ref{bilinear-semi}).
By simple manipulation of eqs.~(\ref{bilinear-semi}), we obtain
$$
\left\{\begin{array}{l}\displaystyle
-2\left(\frac{1}{a}D_t-1\right)f_{k+1}\cdot f_k=g_{k+1}h_k+g_kh_{k+1}\,,
\\[5pt]
2acf_{k+1}f_k=(1+ac)g_{k+1}h_k-(1-ac)g_kh_{k+1}\,,
\\[5pt]
-2af_{k+1}f_k=((1+ac)D_t-a)g_{k+1}\cdot h_k-((1-ac)D_t+a)g_k\cdot h_{k+1}\,.
\end{array}\right.
$$
Let
$$
u_k=\frac{g_k}{f_k}\,,
\qquad
v_k=\frac{h_k}{f_k}\,.
$$
Then we have
\begin{equation}
\left\{\begin{array}{l}\displaystyle
-2\left(\frac{1}{a}\left(\log\frac{f_{k+1}}{f_k}\right)_t-1\right)
=u_{k+1}v_k+u_kv_{k+1}\,,
\\[5pt]
2ac=(1+ac)u_{k+1}v_k-(1-ac)u_kv_{k+1}\,,
\\[5pt]\displaystyle
-2a
=\left[(1+ac)\left(\left(\log\frac{u_{k+1}}{v_k}\right)_t
+\left(\log\frac{f_{k+1}}{f_k}\right)_t\right)-a\right]
u_{k+1}v_k
\\[5pt]\displaystyle
\phantom{-2a}
-\left[(1-ac)\left(\left(\log\frac{u_k}{v_{k+1}}\right)_t-
\left(\log\frac{f_{k+1}}{f_k}\right)_t\right)+a\right]
u_kv_{k+1}\,.
\end{array}\right.\label{dch-derive-a}
\end{equation}
Substituting the first equation into the third equation in
eqs.(\ref{dch-derive-a}), we obtain
$$
\left\{\begin{array}{l}
2ac=(1+ac)u_{k+1}v_k-(1-ac)u_kv_{k+1}\,,
\\[5pt]\displaystyle
-2a
=(1+ac)u_{k+1}v_k\left(\log\frac{u_{k+1}}{v_k}\right)_t
-(1-ac)u_kv_{k+1}\left(\log\frac{u_k}{v_{k+1}}\right)_t
\\[5pt]\displaystyle
\phantom{-2a}
-\frac{a}{2}(u_{k+1}v_k+u_kv_{k+1})((1+ac)u_{k+1}v_k+(1-ac)u_kv_{k+1})
\\[5pt]
\phantom{-2a}
+a^2c(u_{k+1}v_k-u_kv_{k+1})\,.
\end{array}\right.
$$
%%%%%$$
%%%%%\left\{\begin{array}{l}
%%%%%2ac=(1+ac)u_{k+1}v_k-(1-ac)u_kv_{k+1}
%%%%%\\[5pt]\displaystyle
%%%%%-2a=\frac{1}{2}((1+ac)u_{k+1}v_k+(1-ac)u_kv_{k+1})
%%%%%(\log\frac{u_{k+1}v_{k+1}}{u_kv_k})_t\\[5pt]\displaystyle
%%%%%\phantom{-2a}+\frac{1}{2}((1+ac)u_{k+1}v_k-(1-ac)u_kv_{k+1})
%%%%%(\log\frac{u_{k+1}u_k}{v_{k+1}v_k})_t
%%%%%\\[5pt]\displaystyle\phantom{-2a}
%%%%%-\frac{a}{2}(u_{k+1}v_k+u_kv_{k+1})((1+ac)u_{k+1}v_k+(1-ac)u_kv_{k+1})
%%%%%\\[5pt]\displaystyle\phantom{-2a}
%%%%%+\frac{a}{2}(u_{k+1}v_k-u_kv_{k+1})((1+ac)u_{k+1}v_k-(1-ac)u_kv_{k+1})
%%%%%\end{array}\right.
%%%%%$$
Simplifying the above equations, we have
\begin{equation}
\left\{\begin{array}{l}
2ac=(1+ac)u_{k+1}v_k-(1-ac)u_kv_{k+1}\,,
\\[5pt]\displaystyle
-2a
=\frac{1}{2}((1+ac)u_{k+1}v_k+(1-ac)u_kv_{k+1})
\left(\log\frac{u_{k+1}v_{k+1}}{u_kv_k}\right)_t
\\[5pt]\displaystyle
\phantom{-2a}
+ac\left(\log\frac{u_{k+1}u_k}{v_{k+1}v_k}\right)_t-2au_{k+1}v_{k+1}u_kv_k\,,
\end{array}\right.\label{uveq}
\end{equation}
Let
$$
\phi_k=\frac{u_k}{v_k}=\frac{g_k}{h_k}\,,
\qquad
\rho_k=u_kv_k=\frac{g_kh_k}{f_k^2}\,.
$$
{}From the first equation in (\ref{uveq}), we obtain
$$
\frac{2ac}{u_kv_{k+1}}=(1+ac)\frac{\phi_{k+1}}{\phi_k}-(1-ac)\,,
\qquad
\frac{2ac}{u_{k+1}v_k}=(1+ac)-(1-ac)\frac{\phi_k}{\phi_{k+1}}\,.
$$
Multiplying these two equations, we obtain
$$
\frac{(2ac)^2}{\rho_{k+1}\rho_k}
=\left((1+ac)\frac{\phi_{k+1}}{\phi_k}-(1-ac)\right)
\left((1+ac)-(1-ac)\frac{\phi_k}{\phi_{k+1}}\right)\,.
$$
{}From the second equation of (\ref{uveq}), we obtain
$$
-2a
=\frac{1}{2}((1+ac)u_{k+1}v_k+(1-ac)u_kv_{k+1})
\left(\log\frac{\rho_{k+1}}{\rho_k}\right)_t
+ac\left(\log\phi_{k+1}\phi_k\right)_t-2a\rho_{k+1}\rho_k\,.
$$
Thus we have
$$
\frac{(1+ac)u_{k+1}v_k+(1-ac)u_kv_{k+1}}{4ac}
\left(\log\frac{\rho_{k+1}}{\rho_k}\right)_t
+\frac{1}{2}\left(\log\phi_{k+1}\phi_k\right)_t+\frac{1}{c}
=\frac{1}{c}\rho_{k+1}\rho_k\,.
$$
Let us define a lattice parameter
$$
\delta_k=\frac{4ac}{(1+ac)u_{k+1}v_k+(1-ac)u_kv_{k+1}}\,.
$$
Then
\begin{eqnarray*}
&&\delta_k=2\frac{(1+ac)u_{k+1}v_k-(1-ac)u_kv_{k+1}}
{(1+ac)u_{k+1}v_k+(1-ac)u_kv_{k+1}}
=2\frac{(1+ac)g_{k+1}h_k-(1-ac)g_kh_{k+1}}
{(1+ac)g_{k+1}h_k+(1-ac)g_kh_{k+1}}
\\
&&\phantom{\delta_k}
=2\frac{(1+ac)\phi_{k+1}-(1-ac)\phi_k}{(1+ac)\phi_{k+1}+(1-ac)\phi_k}
=2\frac{\displaystyle\frac{1+ac}{1-ac}\frac{\phi_{k+1}}{\phi_k}-1}
{\displaystyle\frac{1+ac}{1-ac}\frac{\phi_{k+1}}{\phi_k}+1}\,.
\end{eqnarray*}
Thus we have
$$
\frac{\phi_{k+1}}{\phi_k}=\frac{1-ac}{1+ac}
\frac{\displaystyle 1+\frac{\delta_k}{2}}{\displaystyle
 1-\frac{\delta_k}{2}}
\,,
$$
where a lattice parameter $\delta_k$ is a function depending on
$(k,t)$.
The lattice parameter corresponds to
$\displaystyle\frac{\partial X}{\partial
x}=\frac{2c}{\rho}=2c+(\log\phi)_x$ in the continuous case.
At time $t$,
$\displaystyle X=X_0+\sum_{k=0}^{K-1}\delta_k$
is a $x$-coordinate of the $k$-th lattice point.
Thus we have the following system:
\begin{equation}
\left\{\begin{array}{l}\displaystyle
\frac{(2ac)^2}{\rho_{k+1}\rho_k}=
\left((1+ac)\frac{\phi_{k+1}}{\phi_k}-(1-ac)\right)
\left((1+ac)-(1-ac)\frac{\phi_k}{\phi_{k+1}}\right)\,,
\\[5pt]\displaystyle
\frac{1}{\delta_k}\left(\log\frac{\rho_{k+1}}{\rho_k}\right)_t
+\frac{1}{2}(\log\phi_{k+1}\phi_k)_t+\frac{1}{c}
=\frac{1}{c}\rho_{k+1}\rho_k\,,
\\[5pt]\displaystyle
\delta_k=2\frac{\displaystyle\frac{1+ac}{1-ac}\frac{\phi_{k+1}}{\phi_k}-1}
{\displaystyle\frac{1+ac}{1-ac}\frac{\phi_{k+1}}{\phi_k}+1}\,.
\end{array}\right.\label{phirhodelta}
\end{equation}
Let
$$
w_k=(\log\phi_k)_t=\left(\log\frac{g_k}{h_k}\right)_t\,.
$$
{}From the first equation in (\ref{phirhodelta}), we obtain
\begin{eqnarray*}
&&-(\log\rho_{k+1}\rho_k)_t
=\frac{\displaystyle\frac{1+ac}{1-ac}\left(\frac{\phi_{k+1}}{\phi_k}\right)_t}
{\displaystyle\frac{1+ac}{1-ac}\frac{\phi_{k+1}}{\phi_k}-1}
+\frac{\displaystyle-\left(\frac{\phi_k}{\phi_{k+1}}\right)_t}
{\displaystyle\frac{1+ac}{1-ac}-\frac{\phi_k}{\phi_{k+1}}}
\\
&&\phantom{-(\log\rho_{k+1}\rho_k)_t}
=\left(\frac{\displaystyle\frac{1+ac}{1-ac}\frac{\phi_{k+1}}{\phi_k}}
{\displaystyle\frac{1+ac}{1-ac}\frac{\phi_{k+1}}{\phi_k}-1}
+\frac{\displaystyle\frac{\phi_k}{\phi_{k+1}}}
{\displaystyle\frac{1+ac}{1-ac}-\frac{\phi_k}{\phi_{k+1}}}\right)
\left(\log\frac{\phi_{k+1}}{\phi_k}\right)_t
\\
&&\phantom{-(\log\rho_{k+1}\rho_k)_t}
=\frac{\displaystyle\frac{1+ac}{1-ac}\frac{\phi_{k+1}}{\phi_k}+1}
{\displaystyle\frac{1+ac}{1-ac}\frac{\phi_{k+1}}{\phi_k}-1}
\left(\log\frac{\phi_{k+1}}{\phi_k}\right)_t
\\
&&\phantom{-(\log\rho_{k+1}\rho_k)_t}
=\frac{2}{\delta_k}(w_{k+1}-w_k)\,.
\end{eqnarray*}
{}From the third equation in (\ref{phirhodelta}), we obtain
$$
\frac{2}{\delta_k}+1
=\frac{\displaystyle 2\frac{1+ac}{1-ac}\frac{\phi_{k+1}}{\phi_k}}
{\displaystyle\frac{1+ac}{1-ac}\frac{\phi_{k+1}}{\phi_k}-1}\,,
\qquad
\frac{2}{\delta_k}-1
=\frac{2}{\displaystyle\frac{1+ac}{1-ac}\frac{\phi_{k+1}}{\phi_k}-1}\,.
$$
Multiplying these two equations, we obtain
$$
\frac{4}{\delta_k^2}-1
=\frac{4}
{\displaystyle
\left(\frac{1+ac}{1-ac}\frac{\phi_{k+1}}{\phi_k}-1\right)
\left(1-\frac{1-ac}{1+ac}\frac{\phi_k}{\phi_{k+1}}\right)}
=\frac{4(1-ac)(1+ac)}
{\displaystyle\frac{(2ac)^2}{\rho_{k+1}\rho_k}}
=\left(\frac{1}{a^2c^2}-1\right)\rho_{k+1}\rho_k\,.
$$
Thus we have a system
$$
\left\{\begin{array}{l}\displaystyle
-(\log\rho_{k+1}\rho_k)_t=2\frac{w_{k+1}-w_k}{\delta_k}\,,
\\[5pt]\displaystyle
\frac{1}{\delta_k}\left(\log\frac{\rho_{k+1}}{\rho_k}\right)_t
+\frac{w_{k+1}+w_k}{2}+\frac{1}{c}
=\frac{1}{c}\rho_{k+1}\rho_k\,,
\\[5pt]\displaystyle
\frac{4}{\delta_k^2}-1=\left(\frac{1}{a^2c^2}-1\right)\rho_{k+1}\rho_k\,.
\end{array}\right.
$$
Let
$$
r_k=\log\rho_k=\log\frac{g_kh_k}{f_k^2}\,.
$$
Then we have
$$
\left\{\begin{array}{l}\displaystyle
-\partial_t(r_{k+1}+r_k)=2\frac{w_{k+1}-w_k}{\delta_k}\,,
\\[5pt]\displaystyle
\frac{1}{\delta_k}\partial_t(r_{k+1}-r_k)
+\frac{w_{k+1}+w_k}{2}
+\frac{1}{c}\frac{\displaystyle 1-\frac{4a^2c^2}{\delta_k^2}}{1-a^2c^2}=0\,,
\\[5pt]\displaystyle
\frac{4}{\delta_k^2}-1=\left(\frac{1}{a^2c^2}-1\right)e^{r_{k+1}+r_k}\,.
\end{array}\right.
$$
Let
$$
r_k'=\partial_tr_k\,.
$$
Then
$$
\left\{\begin{array}{l}\displaystyle
-(r_{k+1}'+r_k')=2\frac{w_{k+1}-w_k}{\delta_k}\,,
\\[5pt]\displaystyle
\frac{1}{\delta_k}(r_{k+1}'-r_k')
+\frac{w_{k+1}+w_k}{2}
+\frac{1}{c}\frac{\displaystyle 1-\frac{4a^2c^2}{\delta_k^2}}{1-a^2c^2}=0\,,
\\[5pt]\displaystyle
\partial_t\delta_k=\left(1-\frac{\delta_k^2}{4}\right)(w_{k+1}-w_k)\,.
\end{array}\right.
$$
Eliminating $r_k'$, we have
\begin{equation}
\left\{\begin{array}{l}\displaystyle
-2\left(\frac{w_{k+1}-w_k}{\delta_k}-\frac{w_k-w_{k-1}}{\delta_{k-1}}\right)
+\delta_k\frac{w_{k+1}+w_k}{2}
+\frac{\delta_k}{c}
\frac{\displaystyle 1-\frac{4a^2c^2}{\delta_k^2}}{1-a^2c^2}
\\[5pt]\displaystyle
\hskip100pt
+\delta_{k-1}\frac{w_k+w_{k-1}}{2}
+\frac{\delta_{k-1}}{c}
\frac{\displaystyle 1-\frac{4a^2c^2}{\delta_{k-1}^2}}{1-a^2c^2}=0\,,
\\[5pt]\displaystyle
\partial_t\delta_k=\left(1-\frac{\delta_k^2}{4}\right)(w_{k+1}-w_k)\,.
\end{array}\right.\label{d-ch-final}
\end{equation}
This is a semi-discrete Camassa-Holm equation.
Note that the lattice parameter depends on the time and space.
\end{proof}

Differentiating the first equation in (\ref{d-ch-final}) with
respect to
$t$, we
obtain
\begin{eqnarray*}
&&-2\left(\frac{w_{k+1}-w_k}{\delta_k}-\frac{w_k-w_{k-1}}{\delta_{k-1}}\right)_t
+\delta_k\frac{\partial_tw_{k+1}+\partial_tw_k}{2}
+\delta_{k-1}\frac{\partial_tw_k+\partial_tw_{k-1}}{2}
\\
&&\qquad
+\left(1-\frac{\delta_k^2}{4}\right)(w_{k+1}-w_k)
\left(\frac{w_{k+1}+w_k}{2}+\frac{1}{c}
\frac{\displaystyle 1+\frac{4a^2c^2}{\delta_k^2}}{1-a^2c^2}\right)
\\
&&\qquad
+\left(1-\frac{\delta_{k-1}^2}{4}\right)(w_k-w_{k-1})
\left(\frac{w_k+w_{k-1}}{2}+\frac{1}{c}
\frac{\displaystyle 1+\frac{4a^2c^2}{\delta_{k-1}^2}}{1-a^2c^2}\right)=0\,,
\end{eqnarray*}
Taking a continuous limit, this leads to the CH equation.

Then we have the following theorem.
\begin{corollary}
The semi-discrete CH equation has a determinant form of $N$-soliton
 solutions.
\end{corollary}
\begin{proof}
{}From Theorem 3.2 and 3.3, the proof is obvious.
\end{proof}

\section{Peakons, solitons and cuspons
in the semi-discrete CH equation}

\subsection{Peakon limit in the semi-discrete Camassa-Holm equation}

It is known that the CH equation has a peakon solution in the
limit $\kappa \to 0$.
Here we consider a peakon limit in the semi-discrete CH equation
(\ref{semi-d-ch}).

In the semi-discrete CH equation (\ref{semi-d-ch}), consider the limit
\[
\kappa \to 0,\quad \frac{\kappa^2}{\delta_k}\to {\bf\delta} (k)\,,
\]
where
${\bf\delta} (k)$ is a dirac delta function.
In this limit, the first equation of the semi-discrete CH equation
(\ref{semi-d-ch}) leads to
$$
\partial_X^2 w-w={\bf \delta}(X-cT)\,.
$$
This differential equation has a solution
\[
w=\sum_{i=1}^NA_i(T)\exp|X-cT|\,,
\]
which is a form of the peakon solution.

Thus if we consider very small $\kappa$ with very small $\delta_k$ at
some points $k$, the solutions of the
semi-discrete CH equation tend to the peakon solutions of the CH
equation.

\subsection{One soliton/cuspon solution}
{}From the determinant formula with $a_{i,1}/a_{i,2}=\pm 1$,
the $\tau$-functions for one soliton/cuspon solution are
\begin{equation}
g \propto 1 \pm \left(\frac{c-p}{c+p} \right)  e^{\theta}\,, \qquad
  h \propto 1 \pm \left(\frac{c+p}{c-p}\right) e^{\theta}\,,
\end{equation}
with $\theta=2p(x-vt-x_0)$, $v=1/(c^2-p^2)$ where $c=1/\kappa^2>0$.
This leads to a
solution
\begin{equation}\label{1-soliton-cuspon}
    w(x,t)= \frac{4p^2cv}{(c^2+p^2)\pm (c^2-p^2) \cosh \theta}\,,
\end{equation}
\begin{equation}\label{Hodo1}
   X=2cx + \log\left(\frac{g}{h}\right)\,,\quad T=t\,,
\end{equation}
where the positive case in Eq.(\ref{1-soliton-cuspon}) stands for one smooth
soliton solution when $p < c$,
while the negative case in Eq.(\ref{1-soliton-cuspon})
stands for one-cuspon solution when
$p>c$. Otherwise, the solution is singular.
Thus Eq.(\ref{1-soliton-cuspon}) for nonsingular cases can be expressed to
\begin{equation}
    w(x,t)= \frac{4p^2cv}{(c^2+p^2)+ |c^2-p^2| \cosh \theta}\,.
\end{equation}

Similarly, for the semi-discrete case, we have
\begin{equation}
g_k \propto 1 + \left| \frac{c-p}{c+p} \right| \left( \frac{1+ap}{1-ap}
\right)^k e^{\theta}\,, \qquad
 h_k \propto 1 + \left| \frac{c+p}{c-p} \right| \left( \frac{1+ap}{1-ap}
\right)^k e^{\theta}\,,
\end{equation}
with $\theta=-2pv(t+x_0)$, resulting in a solution of the form
\begin{equation}\label{1-discrete}
    w_k(t)= \frac{4p^2cv}{(c^2+p^2)+|c^2-p^2| \left[ \left( \frac{1+ap}{1-ap}
\right)^{-k}e^{-\theta} +  \left( \frac{1+ap}{1-ap}
\right)^{k}e^{\theta} \right]},
\end{equation}
in conjunction with a transform between an equidistance mesh ($a$) and
a non-equidistance mesh
\begin{equation}
\delta_k=2\frac{(1+ac)g_{k+1}h_k-(1-ac)g_kh_{k+1}}{(1+ac)g_{k+1}h_k+(1-ac)g_kh_{k+1}}.
\end{equation}
Eq. (\ref{1-discrete}) corresponds to 1-soliton solution when $p<c$,
1-cuspon solution when $p>c$.
\subsection{Two soliton/cuspon solutions}
{}From the determinant formula with $a_{i,1}/a_{i,2}=\pm 1$,
the $\tau$-functions for two soliton/cuspon solution are
\begin{equation*}
  g \propto 1 + \left| \frac{c_1-p_1}{c_1+p_1} \right|e^{\theta_1} +
  \left| \frac{c_2-p_2}{c_2+p_2} \right|e^{\theta_2} +
  \left| \frac{(c_1-p_1)(c_2-p_2)}{(c_1+p_1)(c_2+p_2)} \right|
\left( \frac{p_1-p_2}{p_1+p_2}\right)^2
  e^{\theta_1+\theta_2}\,,
\end{equation*}
\begin{equation*}
  h \propto 1 + \left| \frac{c_1+p_1}{c_1-p_1} \right|e^{\theta_1} +
  \left| \frac{c_2+p_2}{c_2-p_2} \right|e^{\theta_2} +
  \left| \frac{(c_1+p_1)(c_2+p_2)}{(c_1-p_1)(c_2-p_2)} \right|
\left( \frac{p_1-p_2}{p_1+p_2}\right)^2
  e^{\theta_1+\theta_2}\,,
\end{equation*}
with $\theta_1=2p_1(x-v_1t-x_{10})$, $\theta_2=2p_2(x-v_2t-x_{20})$,
$v_1=1/(c_1^2-p_1^2)$, $v_2=1/(c_2^2-p_2^2)$. The parametric
solution can be calculated through
\begin{equation}\label{2-soliton-cuspon}
    w(x,t) = \left (\log\frac{g}{h} \right)_t, \qquad X=2cx +
    \log\left(\frac gh\right), \quad T=t\,,
\end{equation}
whose form is complicated and is omitted here. Note that the above
expression includes two-soliton solution ($p_1<c_1$, $p_2<c_2$),
two-cuspon solution ($p_1>c_1$, $p_2>c_2$), or soliton-cuspon
solution ($p_1<c_1$, $p_2>c_2$).

Similarly, for the semi-discrete case, we have
\begin{eqnarray*}
&& \fl   g_k \propto 1 + \left| \frac{c_1-p_1}{c_1+p_1} \right|
\left( \frac{1+ap_1}{1-ap_1}
\right)^ke^{\theta_1} +
  \left| \frac{c_2-p_2}{c_2+p_2} \right|
\left( \frac{1+ap_2}{1-ap_2}
\right)^k
e^{\theta_2}\\
&&\fl \quad  +
\left| \frac{(c_1-p_1)(c_2-p_2)}{(c_1+p_1)(c_2+p_2)} \right|
\left( \frac{p_1-p_2}{p_1+p_2}\right)^2
\left( \frac{1+ap_1}{1-ap_1}
\right)^k
\left( \frac{1+ap_2}{1-ap_2}
\right)^k
  e^{\theta_1+\theta_2}\,,
\end{eqnarray*}
\begin{eqnarray*}
&&\fl
  h_k \propto 1 + \left| \frac{c_1+p_1}{c_1-p_1} \right|
\left( \frac{1+ap_1}{1-ap_1}
\right)^k
e^{\theta_1} +
  \left| \frac{c_2+p_2}{c_2-p_2} \right|
\left( \frac{1+ap_2}{1-ap_2}
\right)^k
e^{\theta_2} \\
&&\fl \quad +
  \left| \frac{(c_1+p_1)(c_2+p_2)}{(c_1-p_1)(c_2-p_2)} \right|
\left( \frac{p_1-p_2}{p_1+p_2}\right)^2
\left( \frac{1+ap_1}{1-ap_1}
\right)^k
\left( \frac{1+ap_2}{1-ap_2}
\right)^k
  e^{\theta_1+\theta_2}\,,
\end{eqnarray*}
with $\theta_1=2p_1(-v_1t-x_{10})$, $\theta_2=2p_2(-v_2t-x_{20})$.
The solution can be calculated through
\begin{equation}
w_k(t) = \left(\log \frac{g_k} {h_k}\right)_t,
\end{equation}
with a transform
\begin{equation}
\delta_k=2\frac{(1+ac)g_{k+1}h_k-(1-ac)g_kh_{k+1}}
{(1+ac)g_{k+1}h_k+(1-ac)g_kh_{k+1}}\,.
\end{equation}
The form is complicated and is omitted here.

\section{Numerical computations}
In this section, several examples will be illustrated 
to show the integrable semi-discretization of the CH equation 
is a powerful scheme for the numerical solutions of the CH equation. 
They include (1) Propagation of
one-cuspon solution; (2) Interaction of two-cuspon solutions; (3)
Head-on collision of soliton-cuspon. In actual computations, the
initial mesh spacing $\delta_k$, is assigned by
\begin{equation}\label{initial_delta}
\delta_k=2\frac{(1+ac)\phi_{k+1}-(1-ac)\phi_k}{(1+ac)\phi_{k+1}+(1-ac)\phi_k}
\,,
\end{equation}
where $\phi_k=g_k/h_k$ is obtainable from the corresponding
determinant solutions. At each time step, from the second equation
in (\ref{semi-d-ch}),
the evolution of $\delta_k$ can be exactly calculated
by
\begin{equation}\label{delta}
\delta^{n+1}_k=2\frac{c^n_k e^{(w^n_{k+1}-w^n_k)}-1}{c^n_k
e^{(w^n_{k+1}-w^n_k)}+1}\,,
\end{equation}
with $c^n_k=(2+\delta^n_k)/(2-\delta^n_k)$. Then, the solution
$w^{n+1}_k$ is easily computed by solving a tridiagonal linear
system based on the first equation of the scheme. Therefore, the
computation cost is less than other traditional numerical methods.
Furthermore, we comment that the mesh spacing $\delta_k$ is 
driven automatically by the solution during the numerical computation. 
Therefore, we would like to call it the self-adaptive method. 

{\bf Example 1: 1-cuspon propagation.} The parameters taken for
the 1-cuspon solution are $p=10.98$, $c=10.0$. The number of grid is
taken as $100$ in an interval of width of $4$ in the $x$-domain,
which implies a mesh size of $h=0.04$. However, through the
hodograph transformation, this corresponds to an interval of width
$74.34$ in the $X$-domain, which implies an average mesh size of
$0.7434$. 
In subsequent examples except for Example 4, 
the grid number is fixed to be 100. 
The time step size is taken as $\Delta t = 0.0004$. Fig.
1(a) shows the initial condition. Figs. 1 (b)-(c) display the numerical
solutions (solid line) and exact solutions (dotted line) at $t=2,
4$, respectively. The $L_\infty$ norm are $0.0365$ at $t=2$, and
$0.0985$ at $t=4$. It is noted that the numerical error is mainly
due to the numerical dispersion. In other words, even after a fairly
long time, the numerical solution of a cuspon preserves its shape
very well except for a phase shift. The reason for the numerical
dispersion is thought due to the explicit method to solve the
equation for the time evolution of mesh size. The detailed analysis
is left for our another paper focusing on numerical solutions of
the CH equation.

\begin{figure}[htbp]
\centerline{
\includegraphics[scale=0.4]{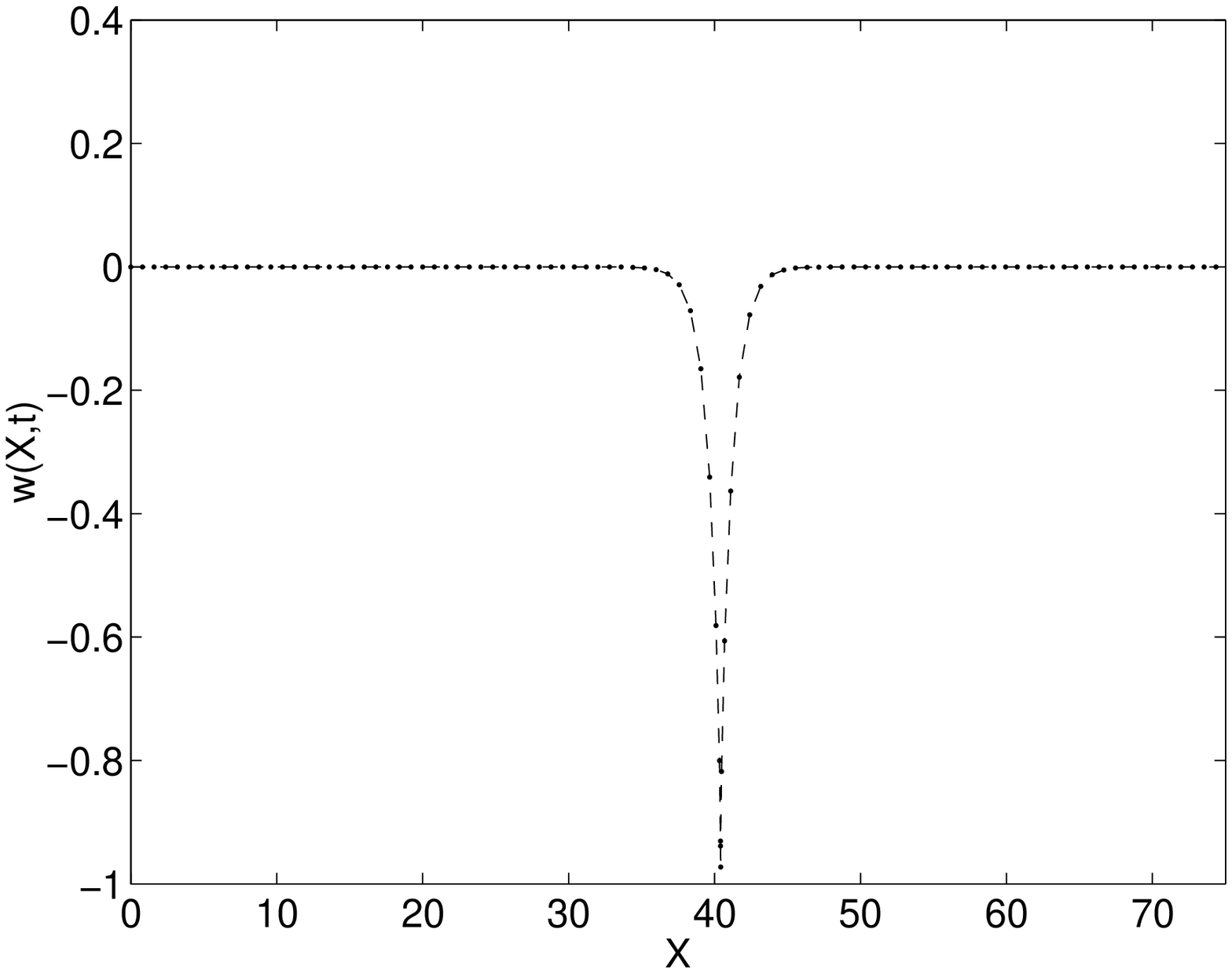}\quad
\includegraphics[scale=0.4]{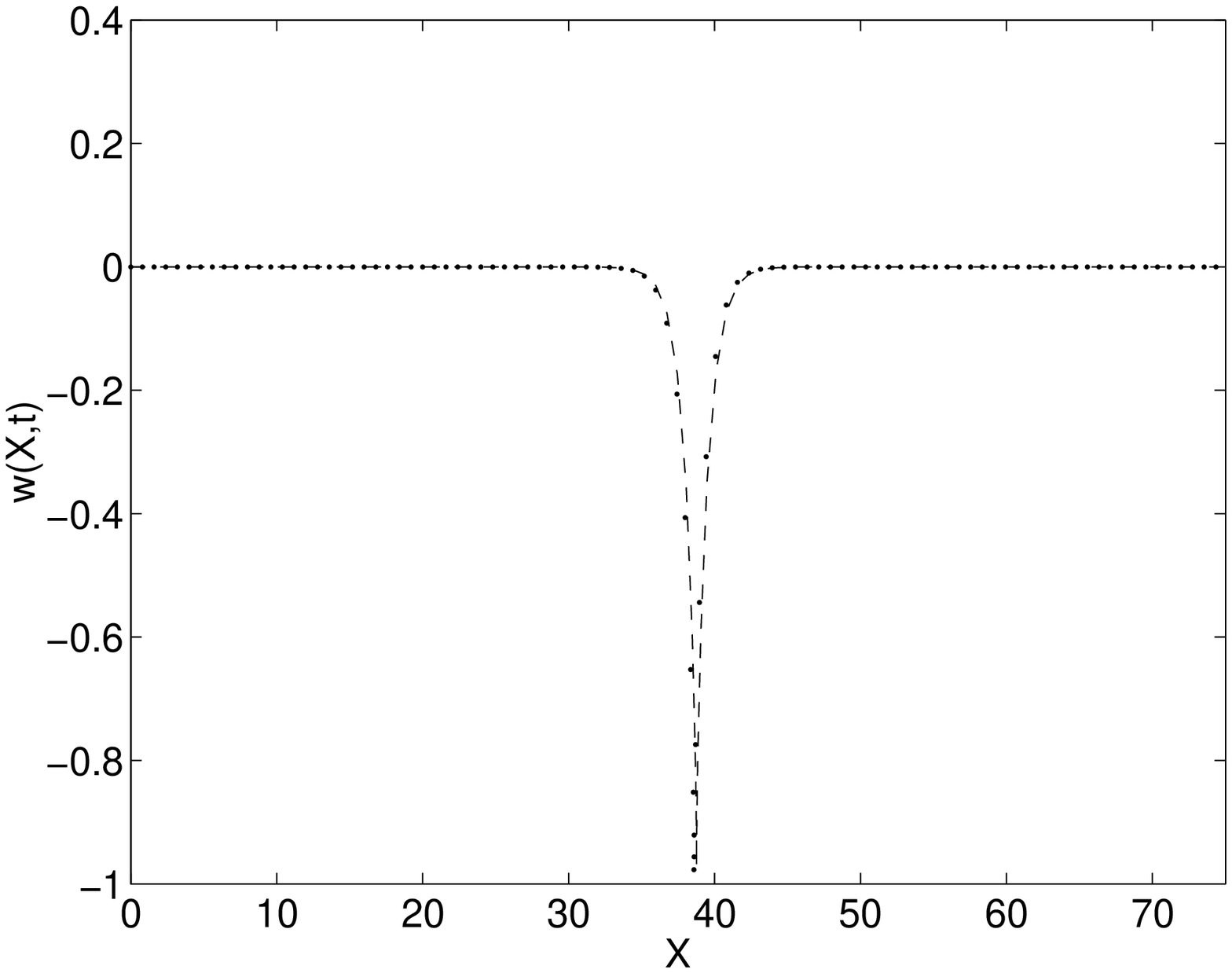}}
\kern-0.355\textwidth \hbox to
\textwidth{\hss(a)\kern0em\hss(b)\kern4em} \kern+0.355\textwidth
\centerline{
\includegraphics[scale=0.4]{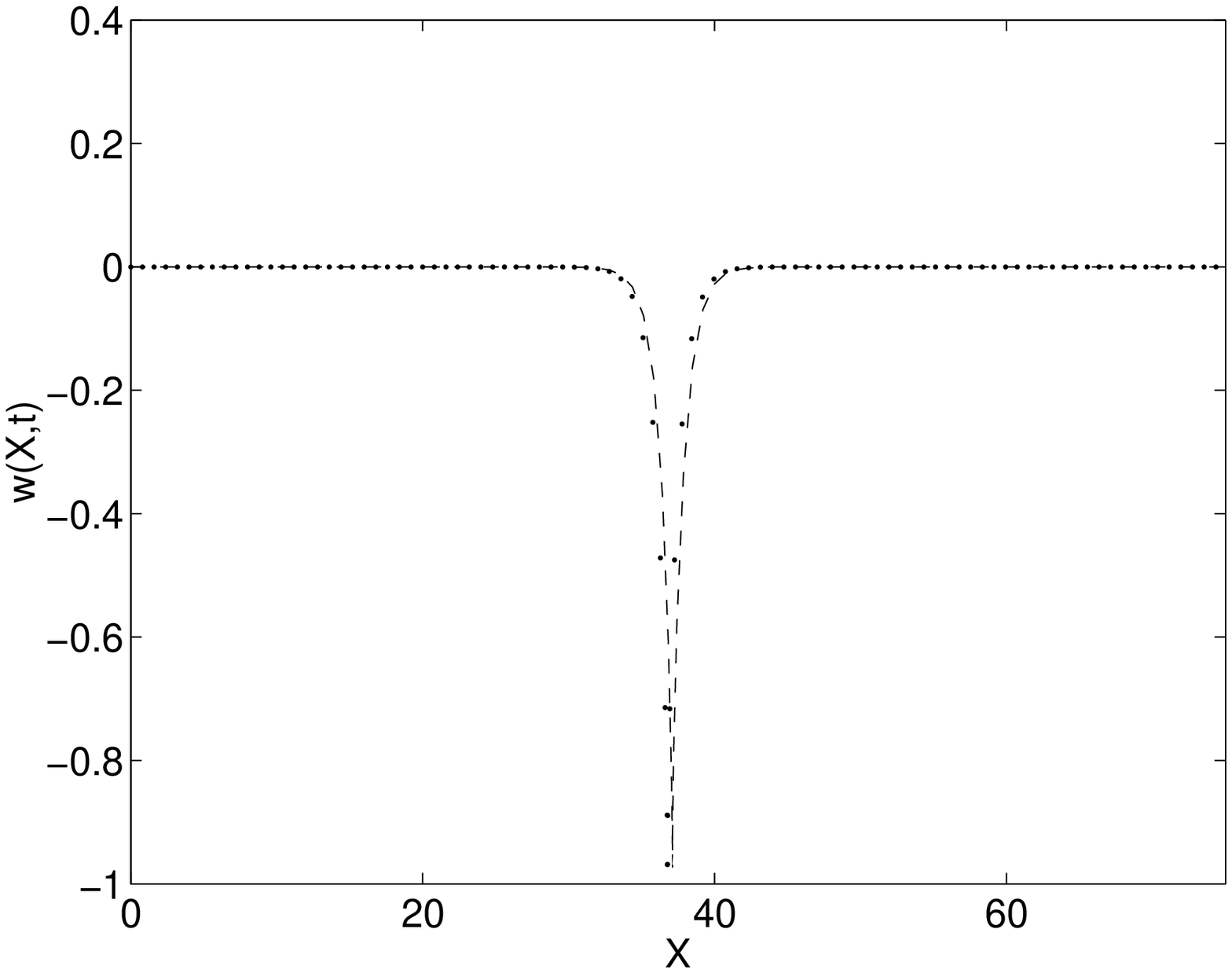}}
\kern-0.355\textwidth \hbox to \textwidth{\hss(c)\kern13em}
\kern+0.355\textwidth
%\kern-0.355\textwidth \hbox to \textwidth{\hss(c)\kern0.0em}
\caption{Numerical solution of one single cuspon solution: (a)
$t=0.0$; (b) $t=2.0$; (c) $t=4.0$.} \label{f:1cuspon}
\end{figure}

{\bf Example 2: 2-cuspon interaction.} The parameters taken for
the two-cuspon solution are $p_1=11.0$, $p_2=10.5$, $c=10.0$.
Fig. 2(a) shows the initial condition, and
Figs. 2(b)-2(e) display the process of
collision at several different times. As far as we know, what is
shown here is the first numerical demonstration for the
cuspon-cuspon interaction due to the singularities of cuspon
solutions. As shown in Fig. 2(e), the 2-cuspon solution regain their
shapes after the collision, only resulting in a phase shift. As
mentioned in \cite{DaiLi}, the two cuspon points are always present
during the collision.

\begin{figure}[htbp]
\centerline{
\includegraphics[scale=0.4]{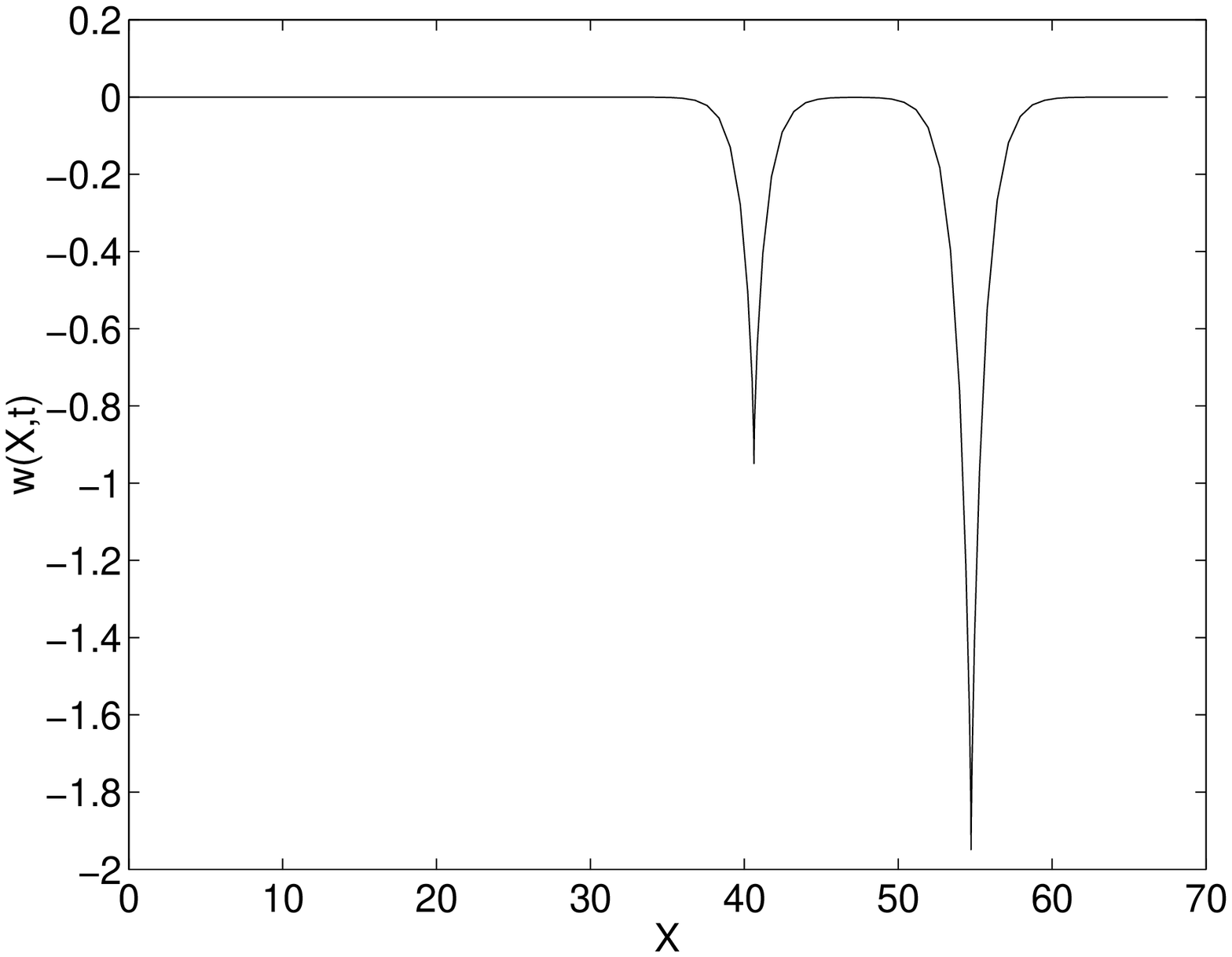}\quad
\includegraphics[scale=0.4]{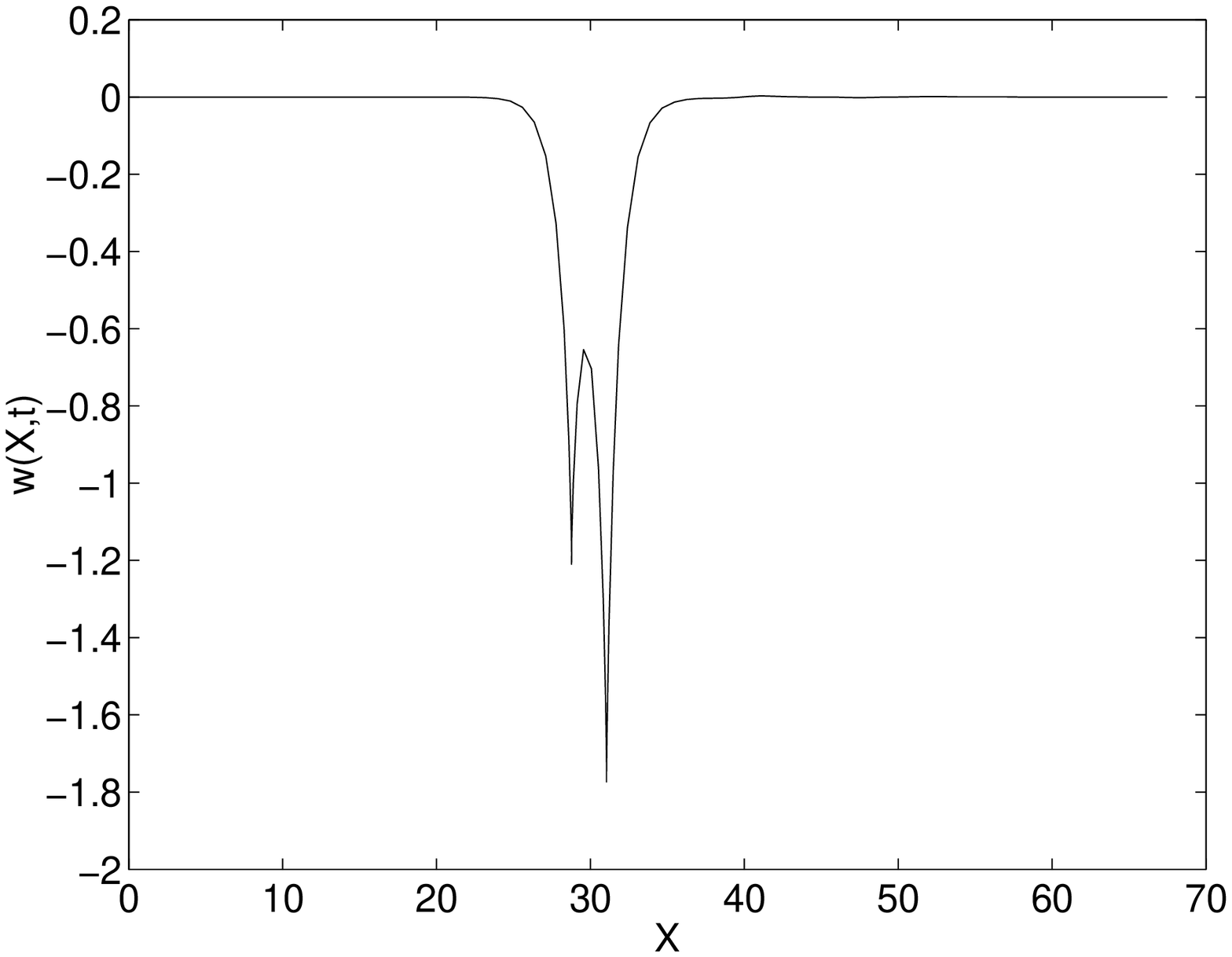}}
\kern-0.1\textwidth \hbox to
\textwidth{\hss(a)\kern0em\hss(b)\kern4em}
\kern+0.1\textwidth\centerline{
\includegraphics[scale=0.4]{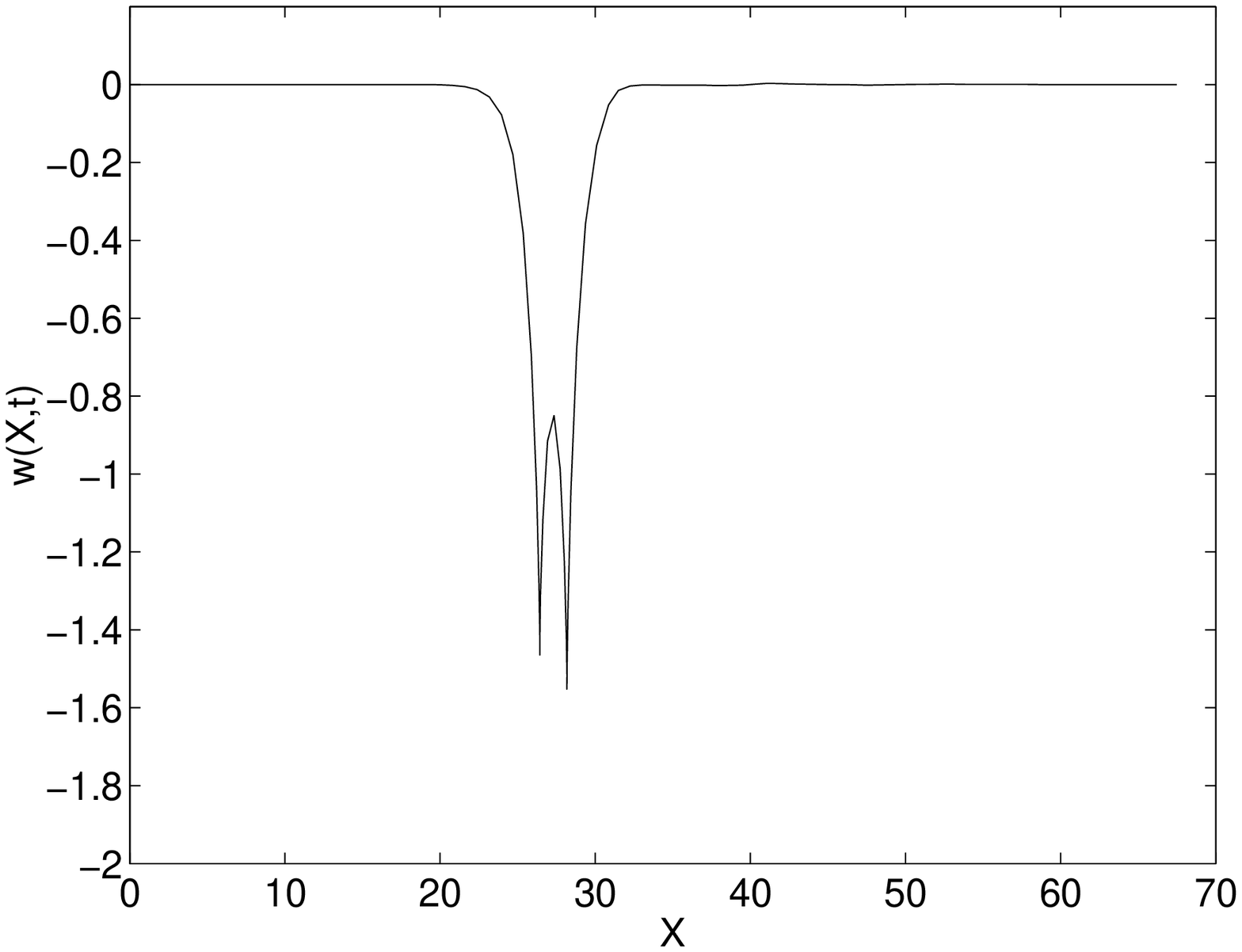}\quad
\includegraphics[scale=0.4]{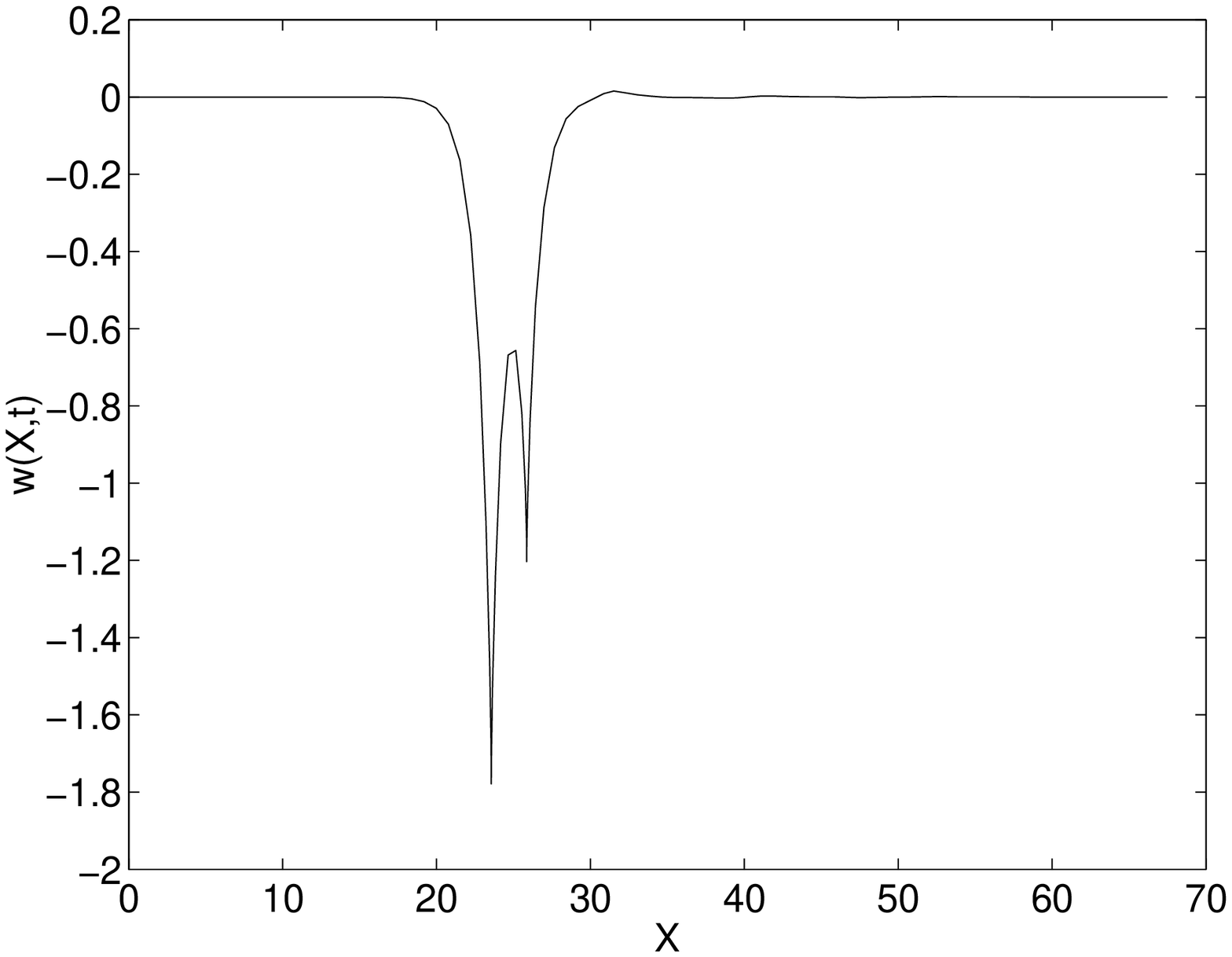}}
\kern-0.1\textwidth \hbox to
\textwidth{\hss(c)\kern0em\hss(d)\kern4em} \kern+0.1\textwidth
\centerline{
\includegraphics[scale=0.4]{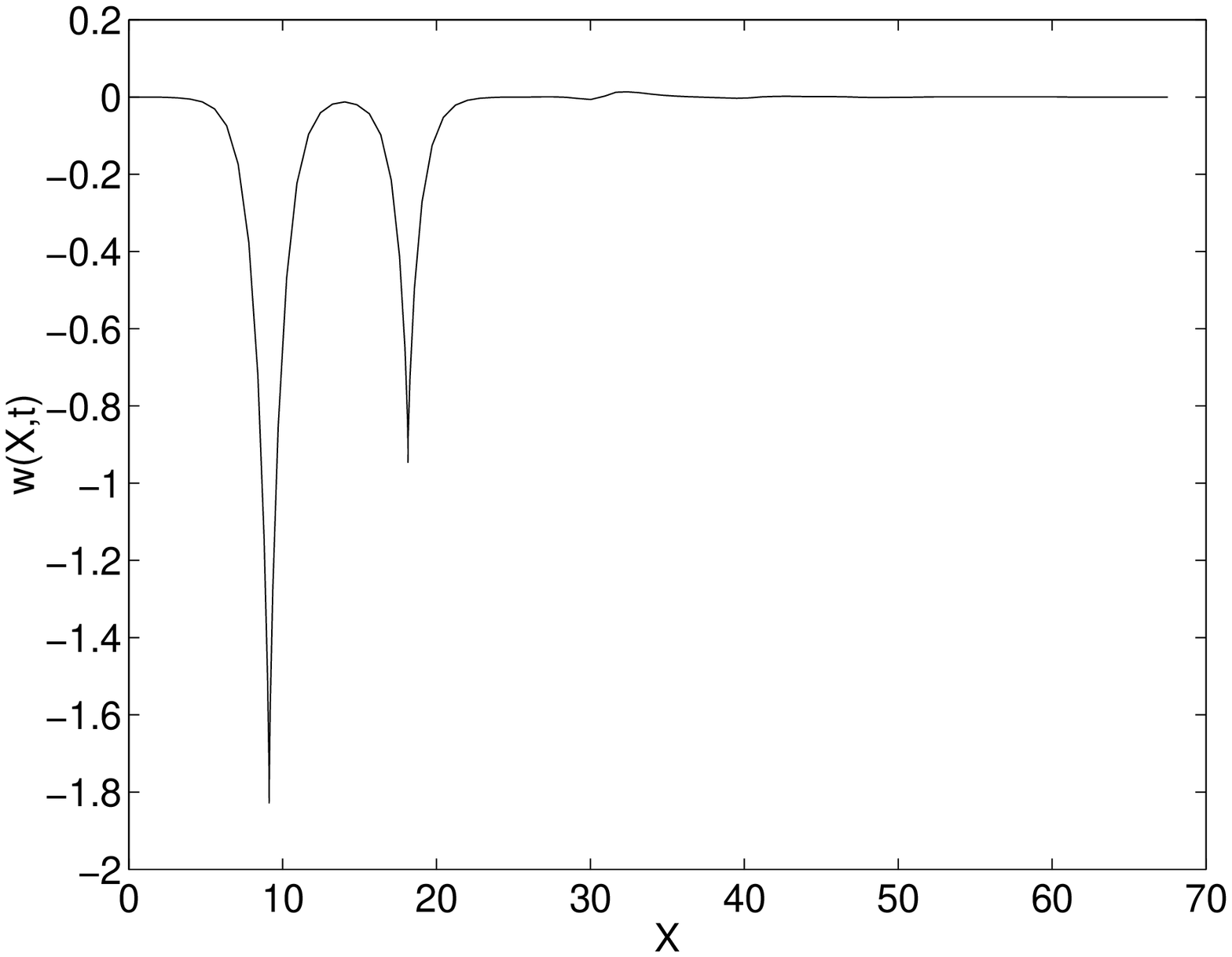}}
\kern-0.1\textwidth \hbox to \textwidth{\hss(e)\kern13em}
\kern+0.1\textwidth
%\kern0.0 \textwidth{\hss(e)\kern0em}
\caption{Numerical solution for
the collision of two-cuspon solution with $p_1=11.0$, $p_2=10.5$,
$c=10.0$: (a) $t=0.0$; (b) $t=13.0$; (c)
 $t=14.8$; (d) $t=16.6$; (e) $t=25.0$.} \label{f:2cuspon}
\end{figure}

{\bf Example 3: Soliton-cuspon interaction.} Here we show two
examples for the soliton-cuspon interaction with $c=10.0$.
In Fig. 3,
we plot the interaction process between a soliton of $p_1=9.12$
and a cuspon of $p_2=10.98$ at several different times where the
soliton and the cuspon have almost the same amplitude. It can be
seen that another singularity
point with infinite derivative ($w_x$) occurs when the
collision starts ($t=12.0$). As collision goes on
($t=14.4, 14.6, 14.8$), the soliton seems 'eats up' the cuspon, and
the profile looks like a complete elevation. However, the cuspon
point is present at all times, especially, at $t=14.6$, the profile
becomes one symmetrical hump with a cuspon point in the middle of
the hump.

\begin{figure}[htbp]
\centerline{
\includegraphics[scale=0.35]{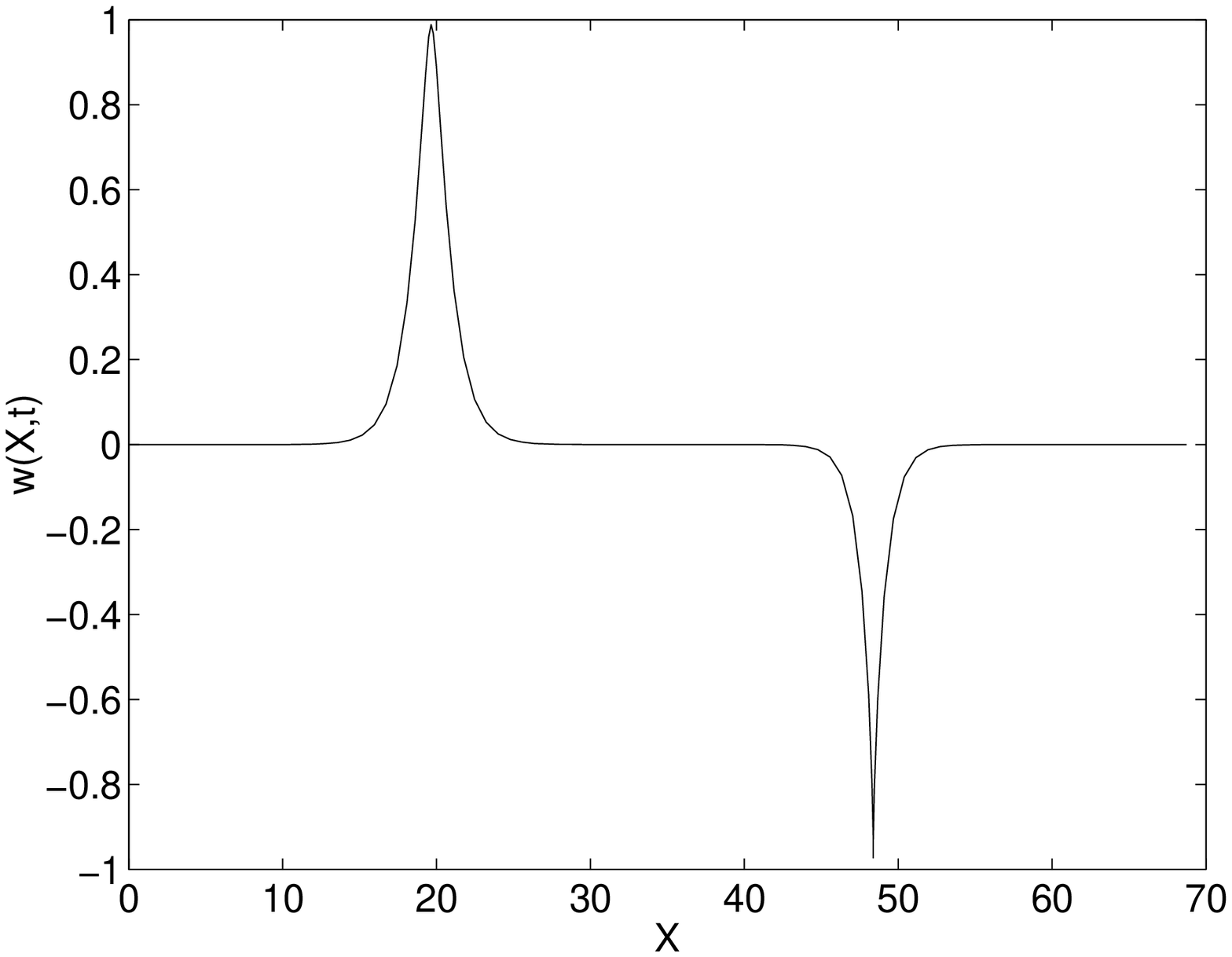}\quad
\includegraphics[scale=0.35]{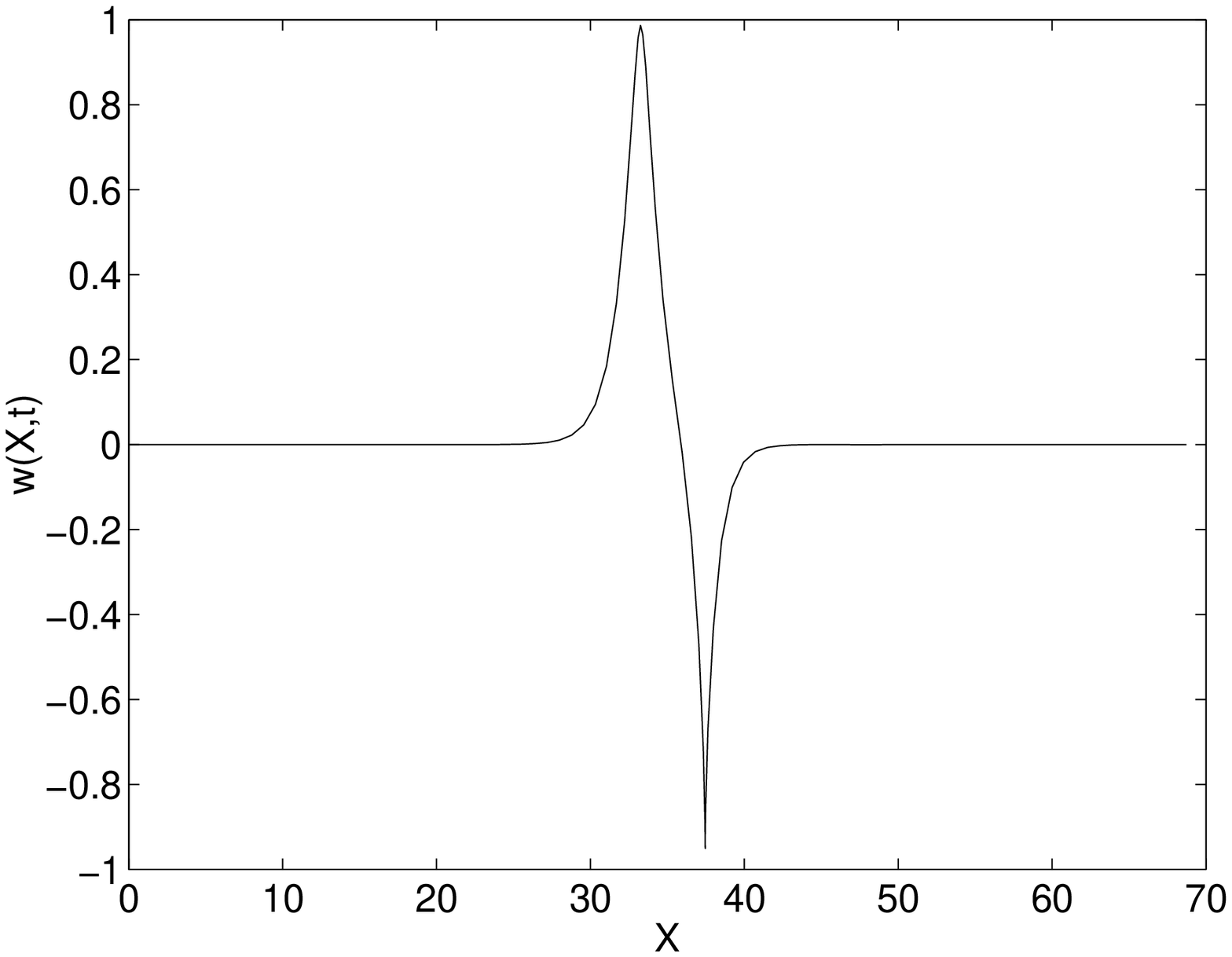}}
\kern-0.315\textwidth \hbox to
\textwidth{\hss(a)\kern0em\hss(b)\kern4em} \kern+0.315\textwidth
\centerline{
\includegraphics[scale=0.35]{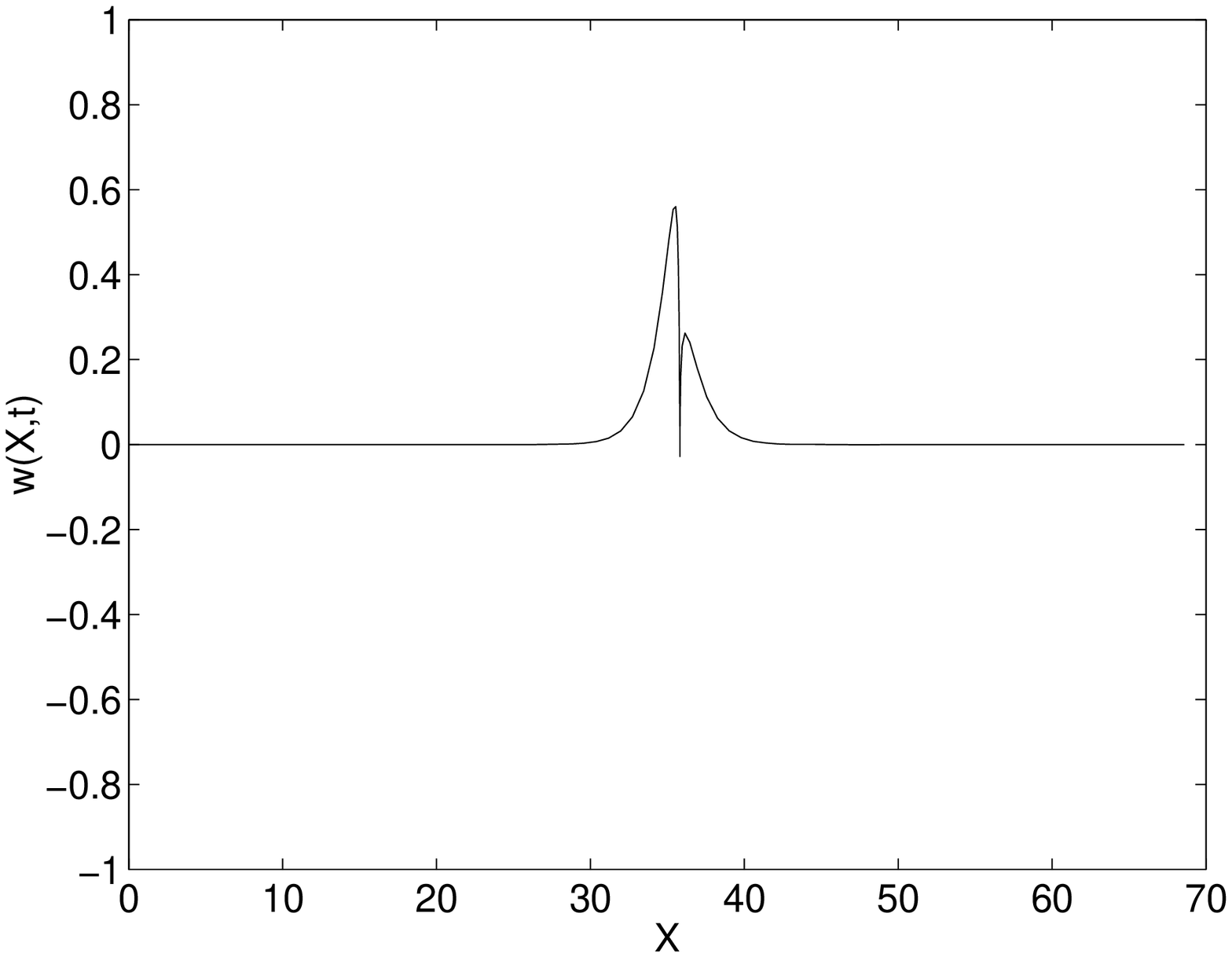}\quad
\includegraphics[scale=0.35]{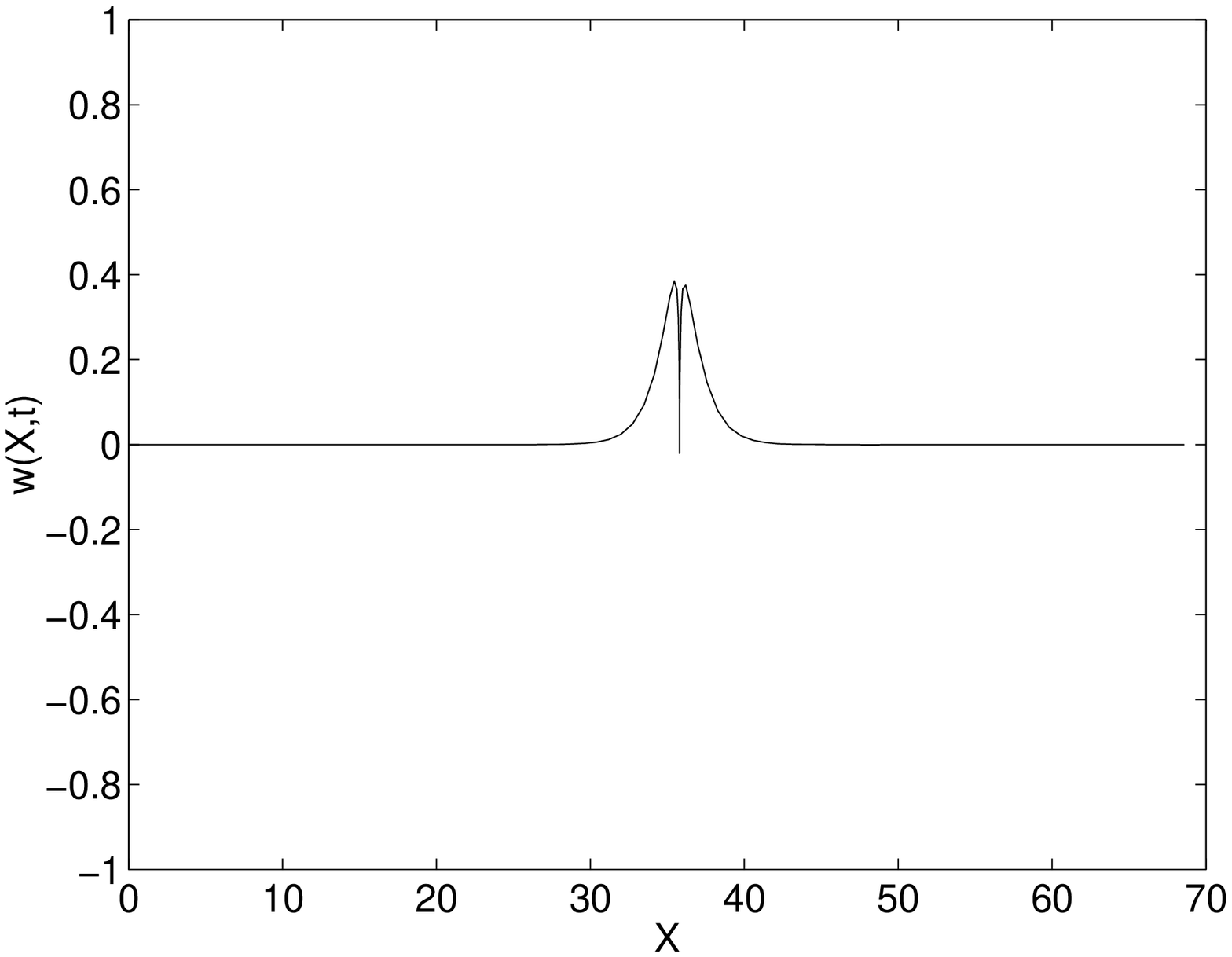}}
\kern-0.315\textwidth \hbox to
\textwidth{\hss(c)\kern0em\hss(d)\kern4em} \kern+0.315\textwidth
\centerline{
\includegraphics[scale=0.35]{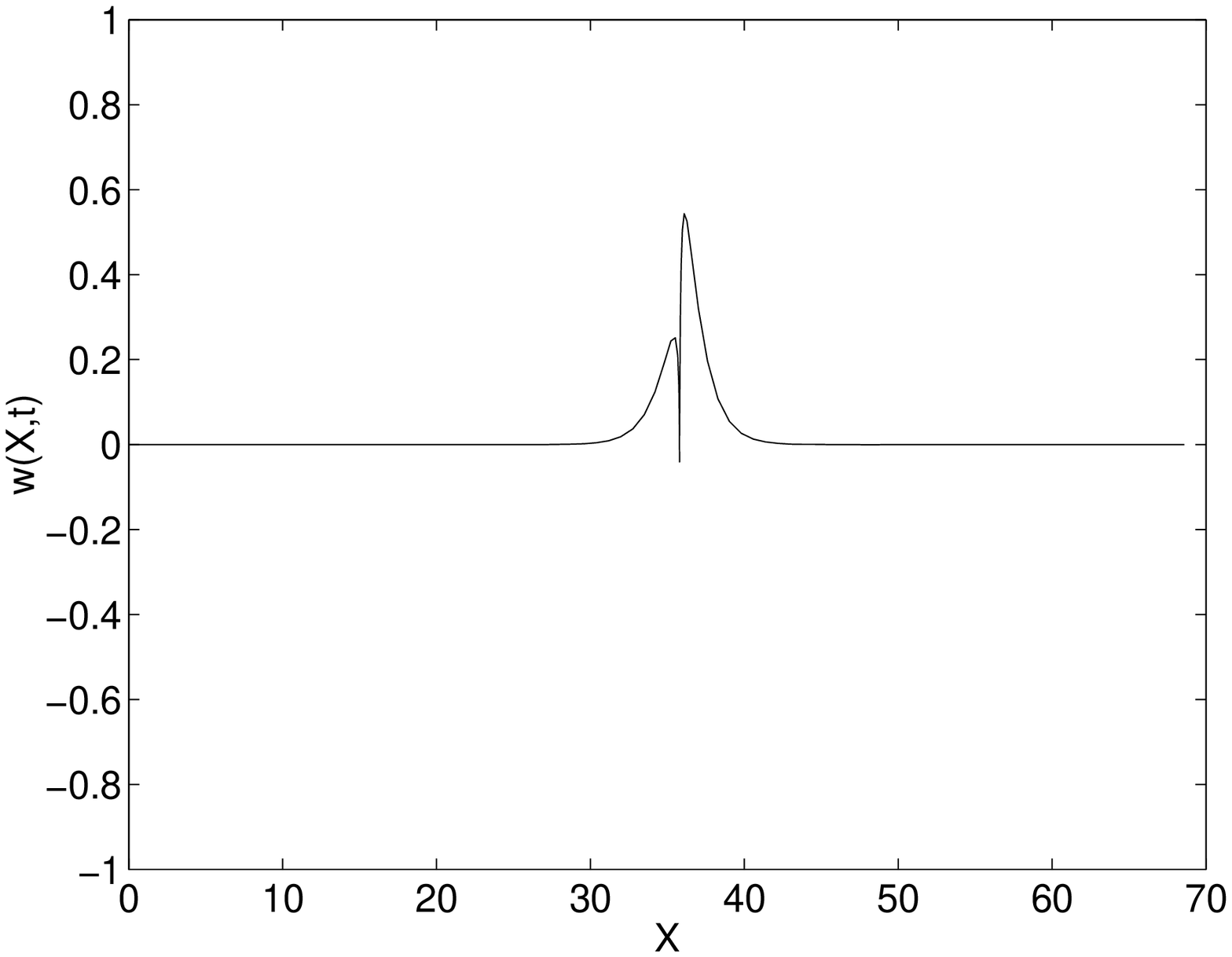}\quad
\includegraphics[scale=0.35]{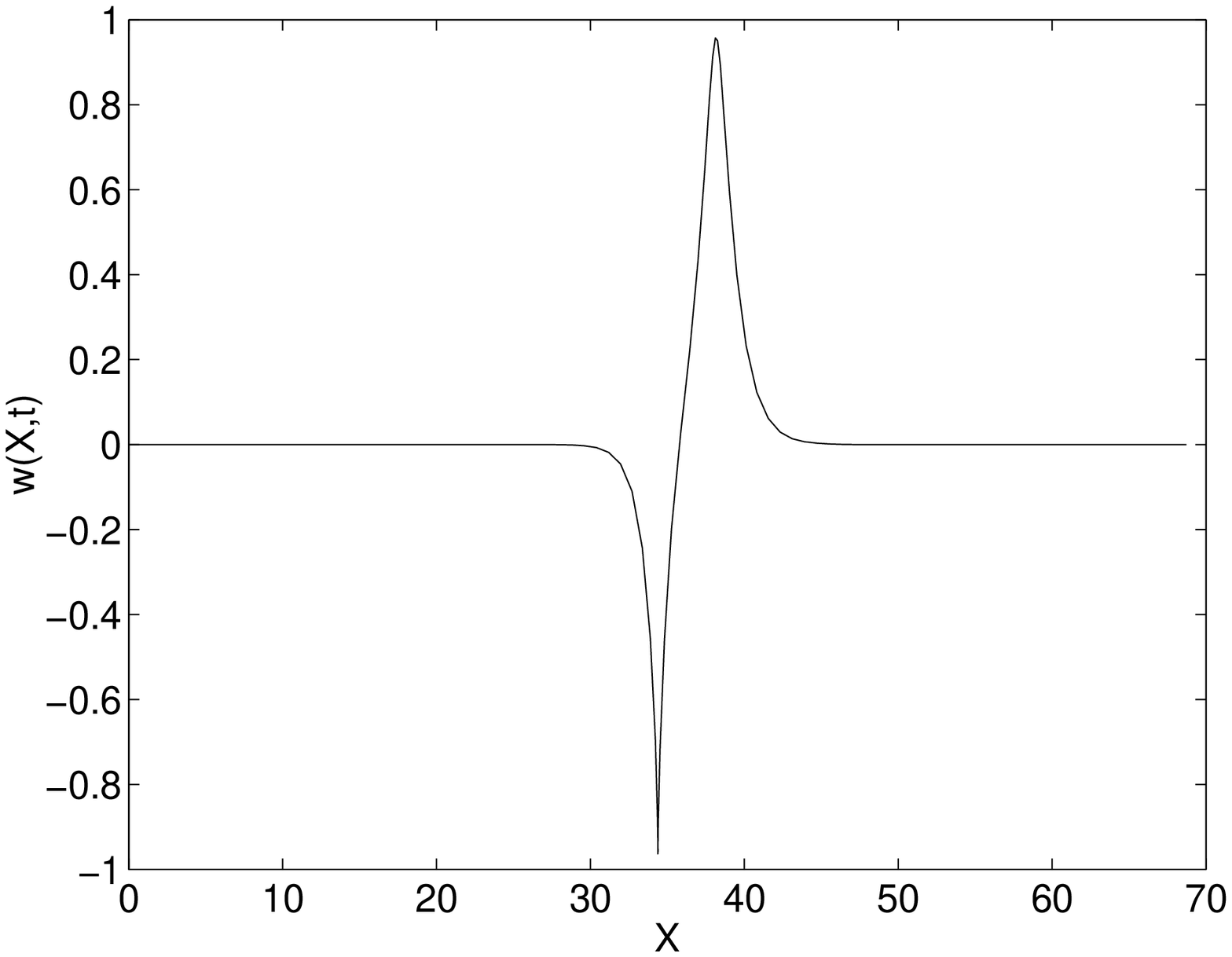}}
\kern-0.315\textwidth \hbox to
\textwidth{\hss(e)\kern0em\hss(f)\kern4em} \kern+0.315\textwidth
\centerline{
\includegraphics[scale=0.35]{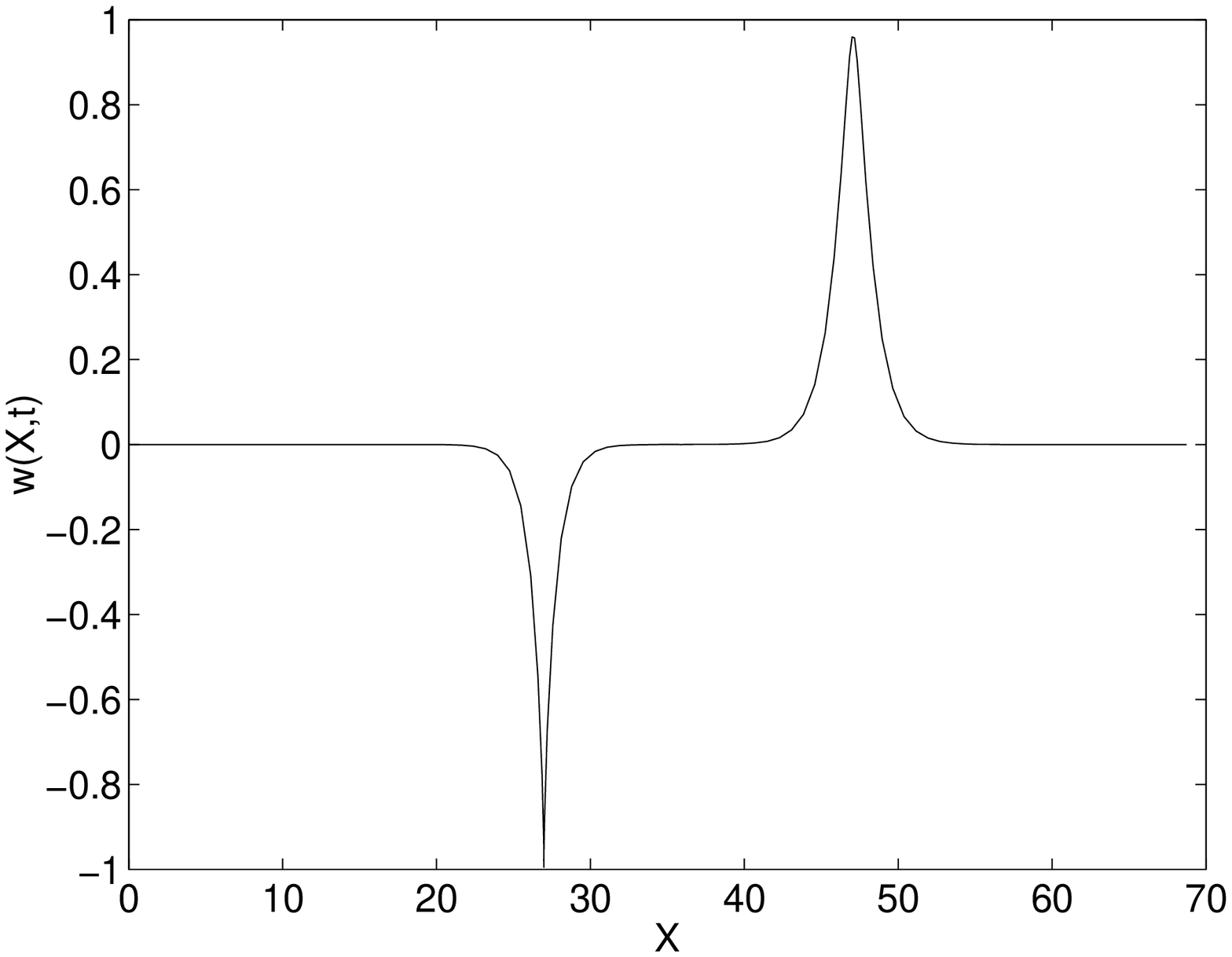}}
\kern-0.315\textwidth \hbox to \textwidth{\hss(g)\kern13em}
\kern+0.315\textwidth
%\quad
%\includegraphics[scale=0.4]{stcpp19p12p210p98t17.eps}}
%\kern-0.0\textwidth{\hss(g)\kern0em}
%\kern+0.355\textwidth
\caption{Numerical solution for soliton-cuspon collision with
$p_1=9.12$, $p_2=10.98$ and $c=10.0$: (a) $t=0.0$; (b) $t=12.0$; (c)
 $t=14.4$; (d) $t=14.6$; (e) $t=14.8$; (f) $t=17.0$; (g) $t=25.0$; }
\label{f:cuspon_soliton}
\end{figure}

In Figure 4, we present another example of a collision between a
soliton ($p_1=9.12$) and a cuspon ($p_2=10.5$) where the cuspon has
a larger amplitude ($2.0$) than the soliton ($1.0$). Again, when the
collision starts, another singularity point appears. As collision
goes on, the soliton is gradually absorbed by the cuspon. At
$t=10.3$, the whole profile looks like a single cuspon when the
soliton is completely absorbed. Later on, the soliton emerges from
the right until $t=16$, the soliton and cuspon recover their
original shapes except for a phase shift when the collision is
complete.

\begin{figure}[htbp]
\centerline{
\includegraphics[scale=0.35]{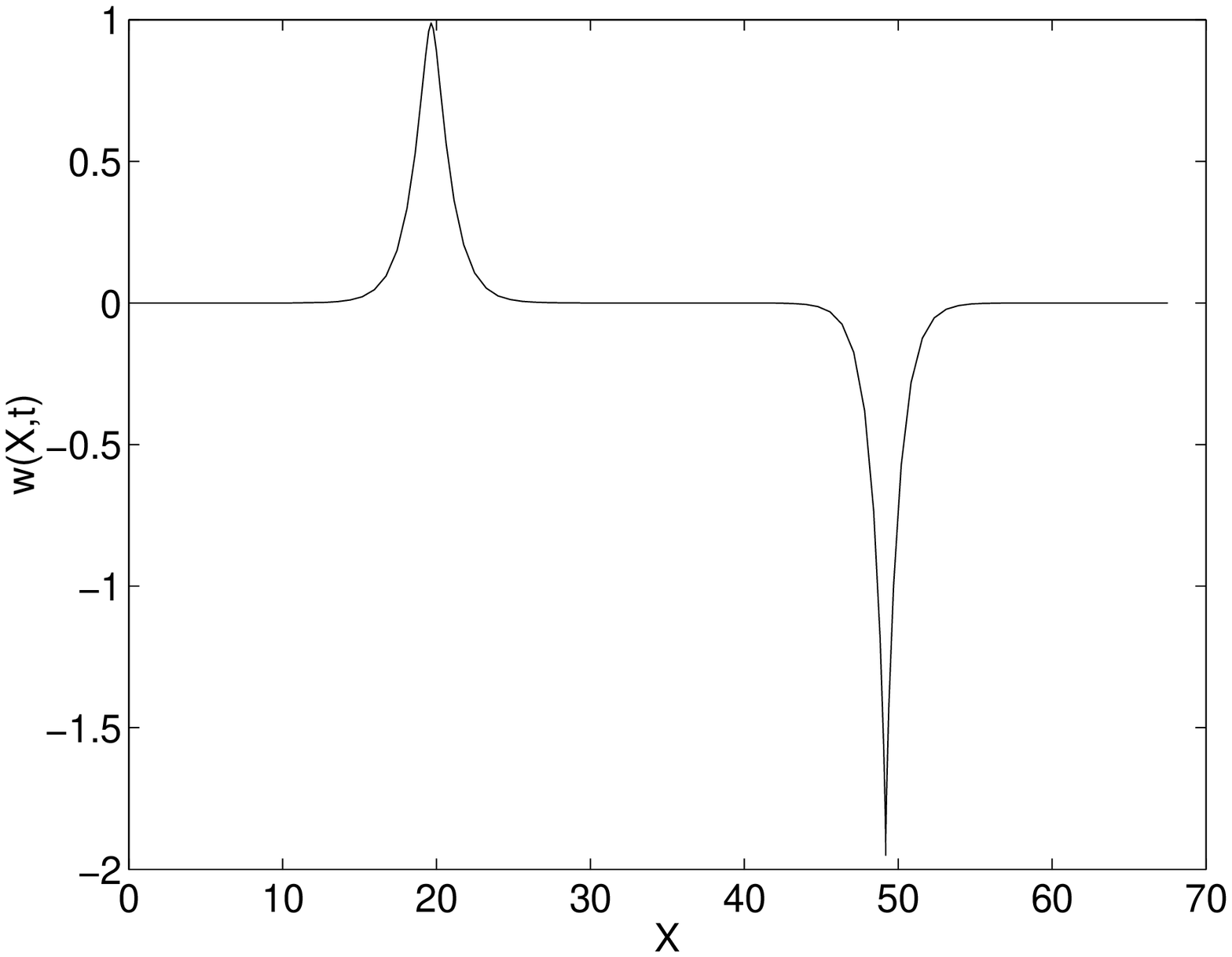}\quad
\includegraphics[scale=0.35]{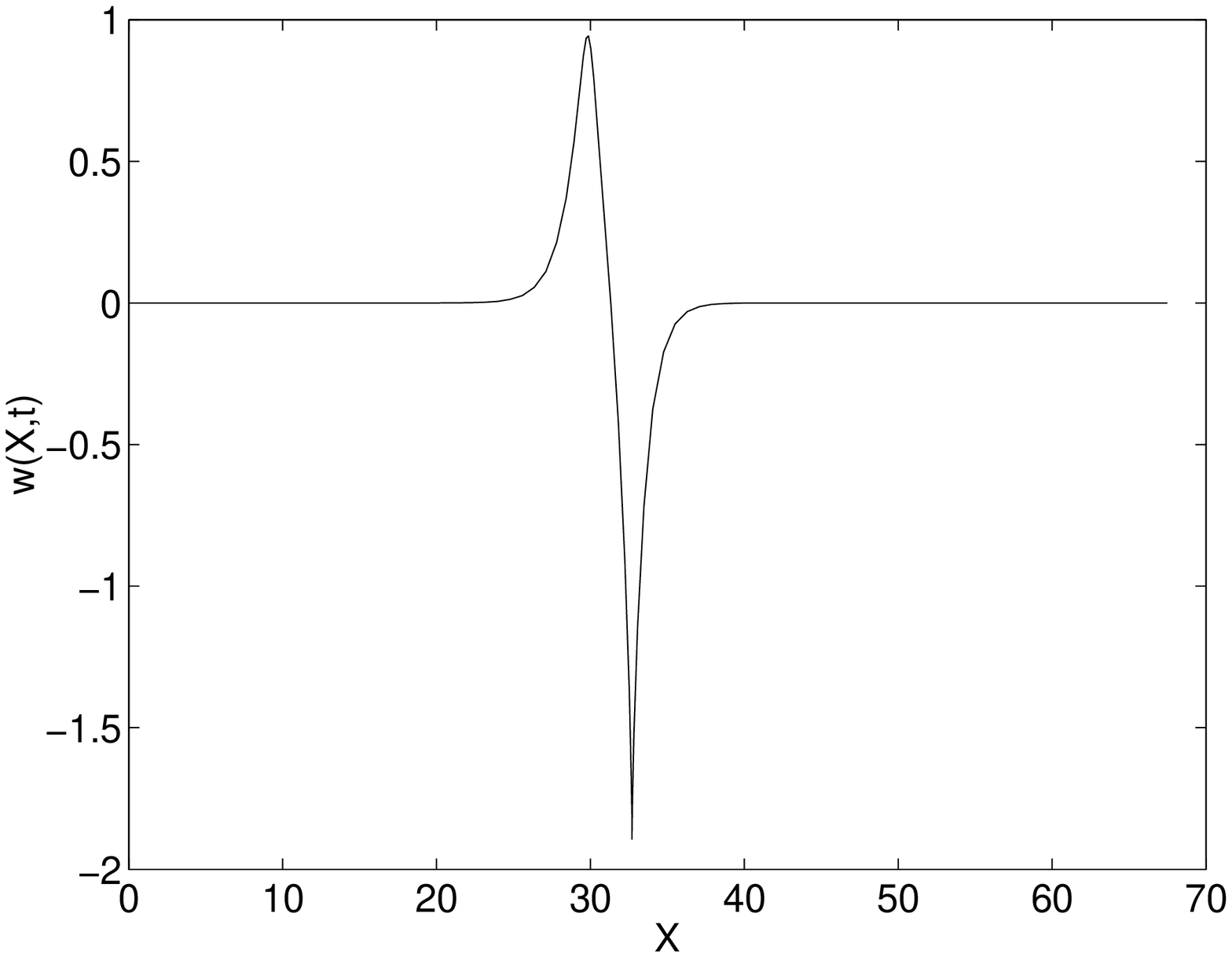}}
\kern-0.315\textwidth \hbox to
\textwidth{\hss(a)\kern0em\hss(b)\kern4em} \kern+0.315\textwidth
\centerline{
\includegraphics[scale=0.35]{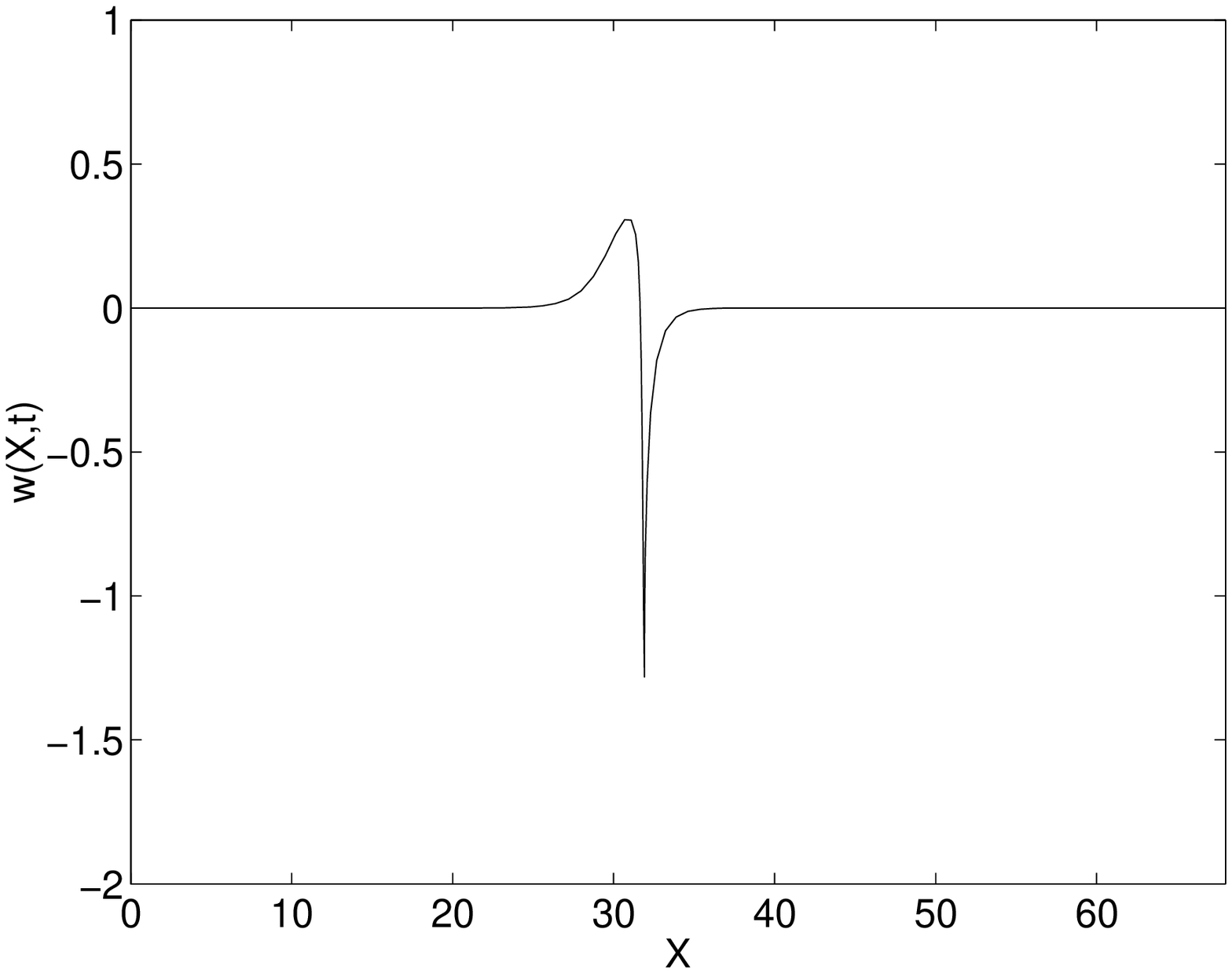}\quad
\includegraphics[scale=0.35]{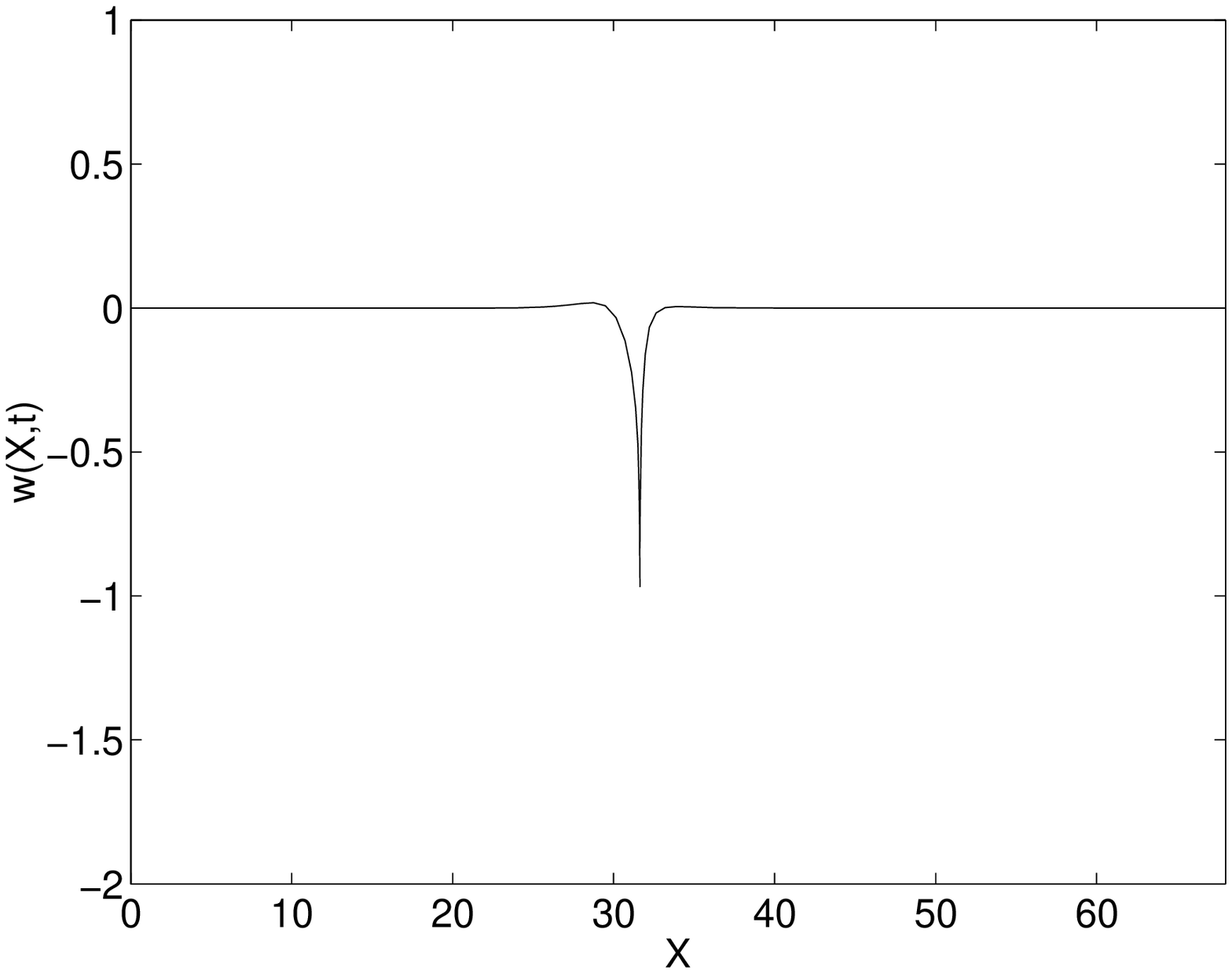}}
\kern-0.315\textwidth \hbox to
\textwidth{\hss(c)\kern0em\hss(d)\kern4em} \kern+0.315\textwidth
\centerline{
\includegraphics[scale=0.35]{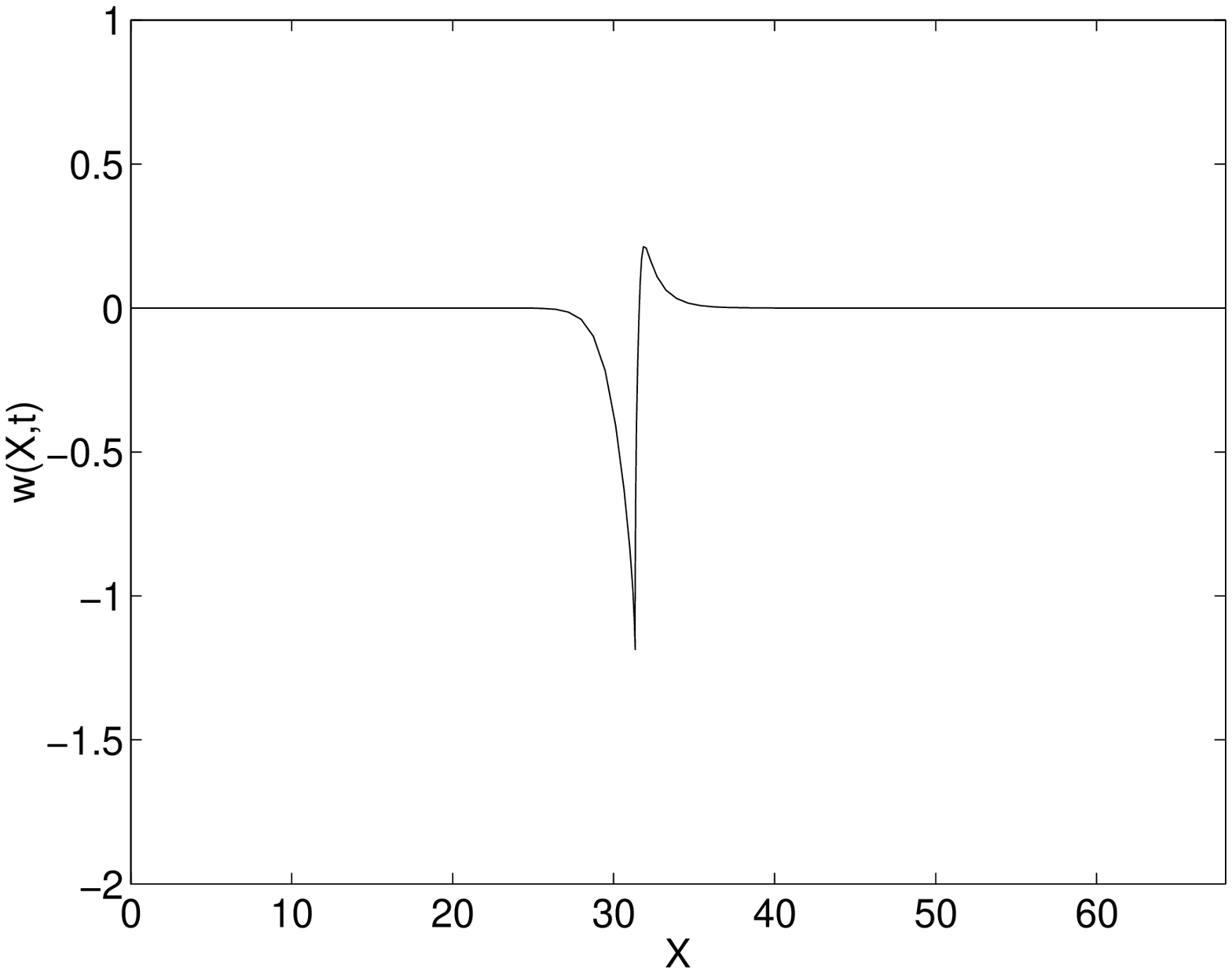}\quad
\includegraphics[scale=0.35]{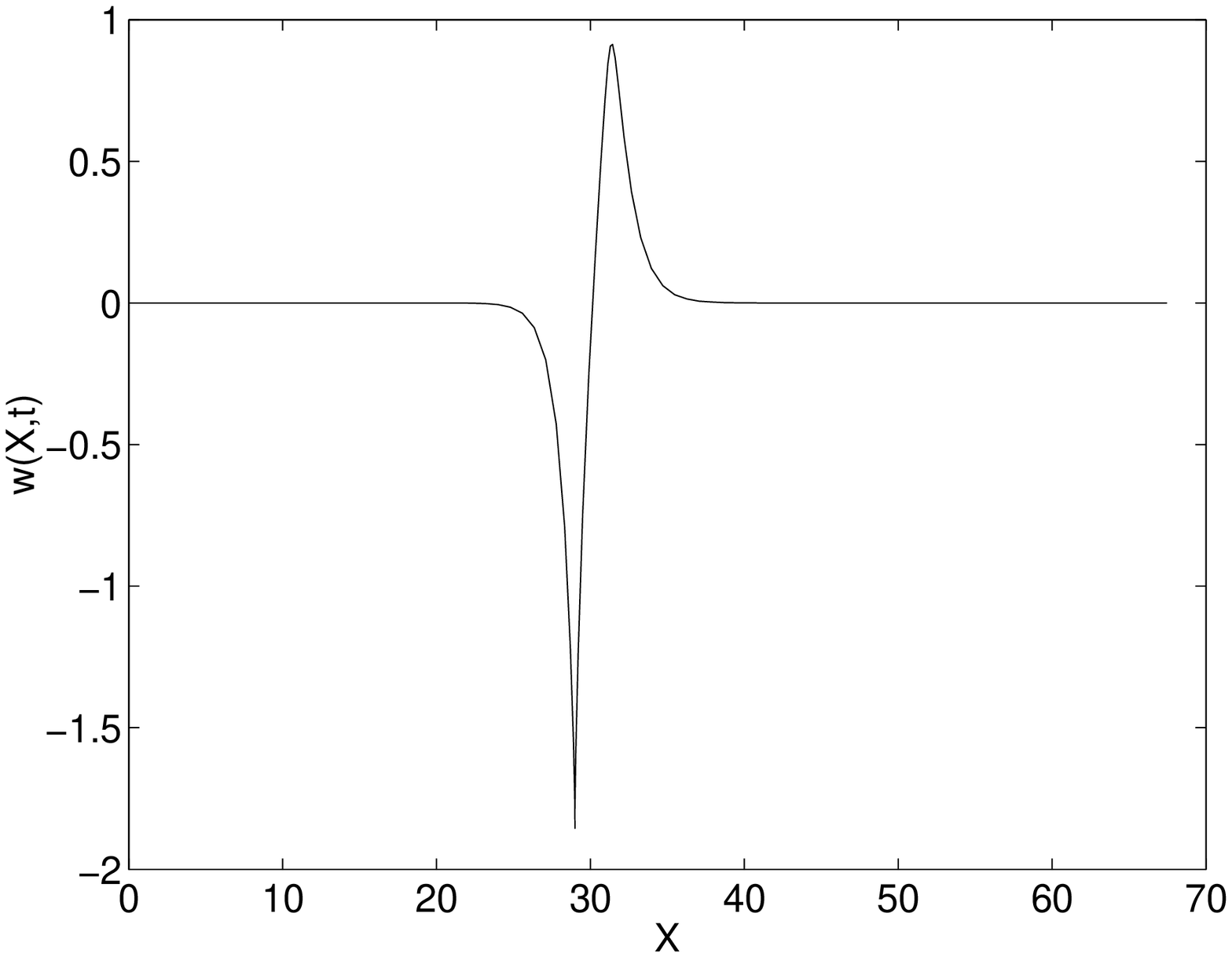}}
\kern-0.315\textwidth \hbox to
\textwidth{\hss(e)\kern0em\hss(f)\kern4em} \kern+0.315\textwidth
\centerline{
\includegraphics[scale=0.35]{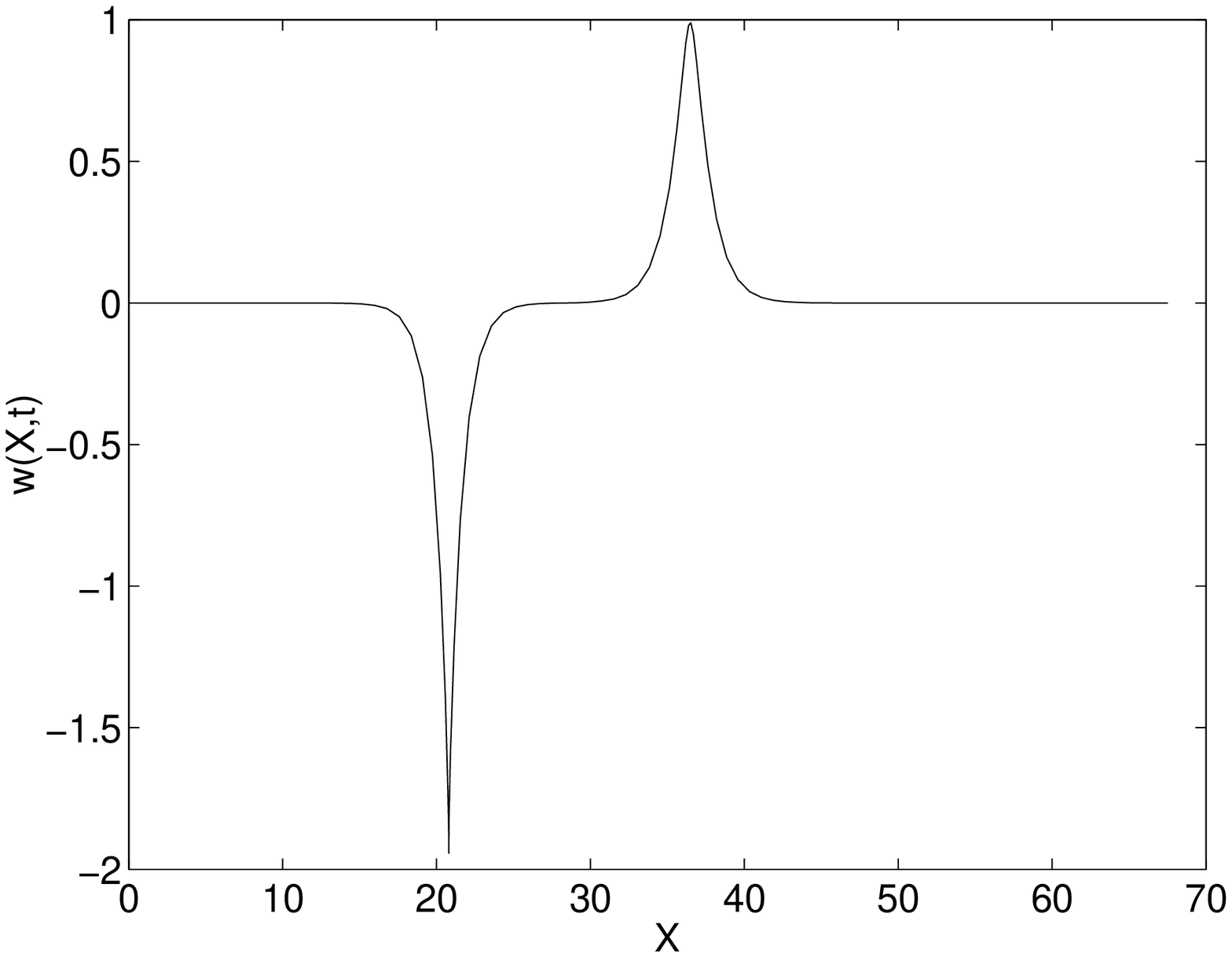}}
\kern-0.315\textwidth \hbox to \textwidth{\hss(g)\kern13em}
\kern+0.315\textwidth
%\quad
%\includegraphics[scale=0.4]{stcpp19p12p210p98t17.eps}}
%\kern-0.0\textwidth{\hss(g)\kern0em}
%\kern+0.355\textwidth
\caption{Numerical solution for soliton-cuspon collision with
$p_1=9.12$, $p_2=10.5$ and $c=10.0$: (a) $t=0.0$; (b) $t=9.0$; (c)
 $t=10.0$; (d) $t=10.3$; (e) $t=10.6$; (f) $t=11.5$; (g) $t=16.0$.}
\label{f:cuspon_soliton2}
\end{figure}

\begin{figure}[htbp]
\centerline{
\includegraphics[scale=0.35]{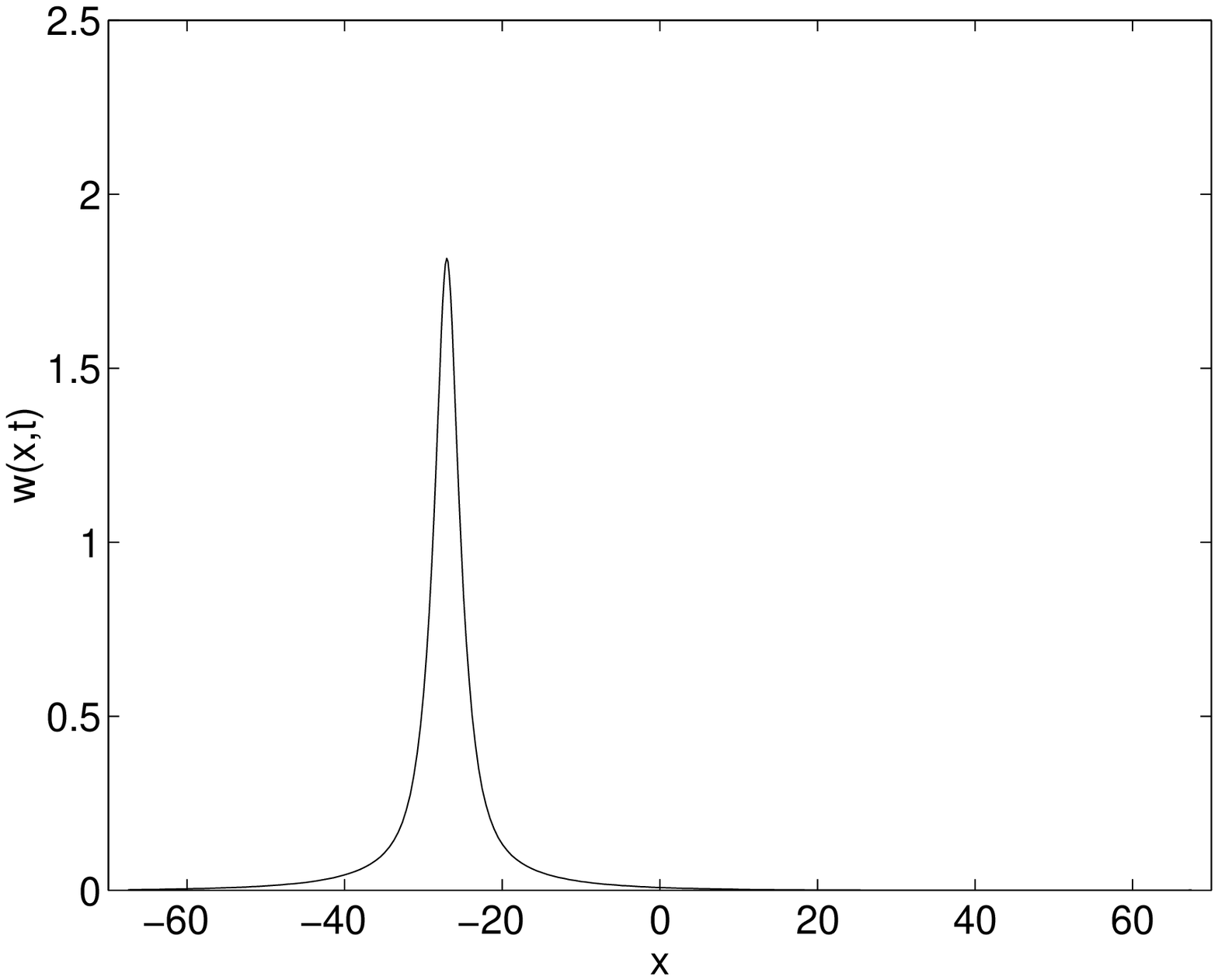}\quad
\includegraphics[scale=0.35]{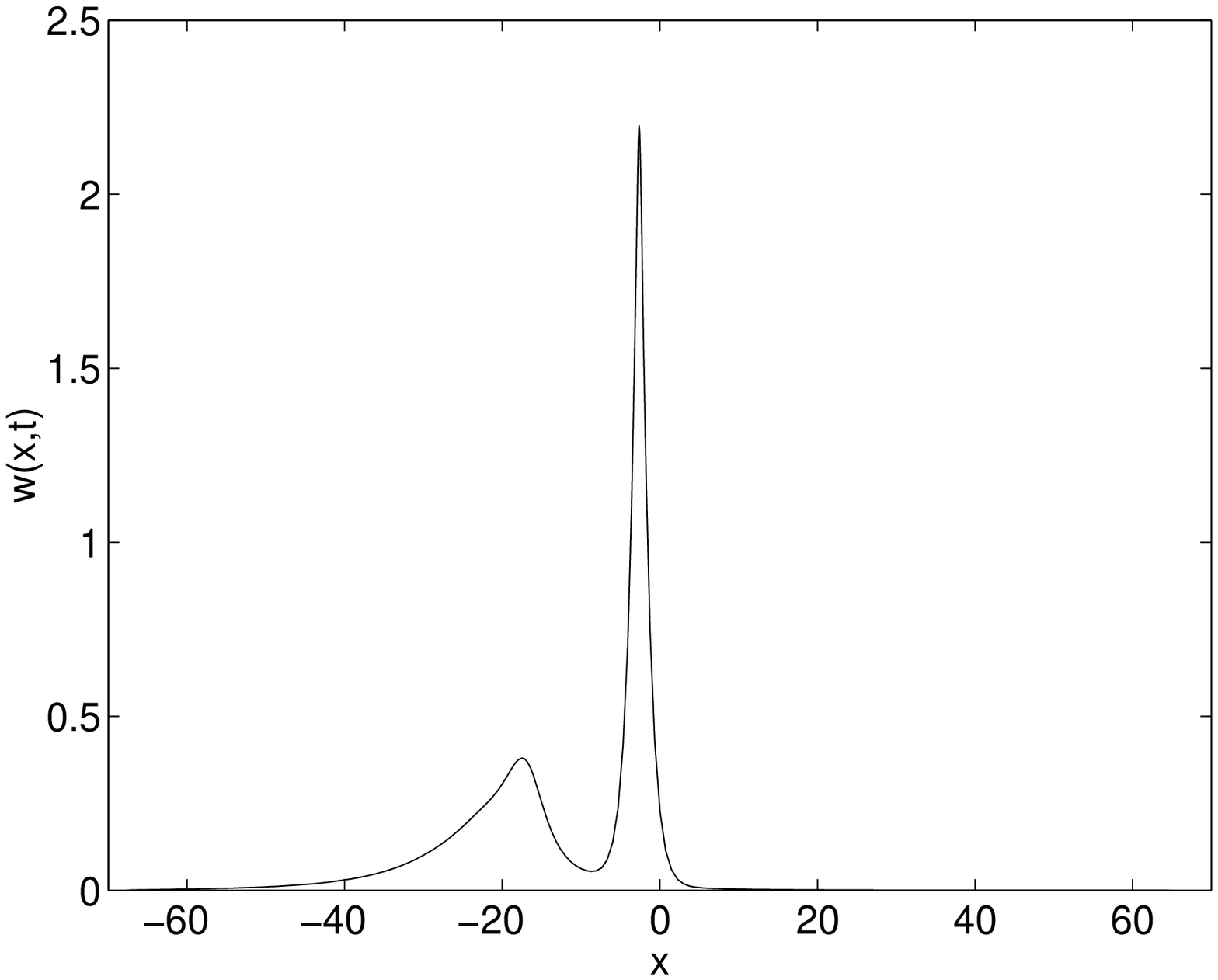}}
\kern-0.315\textwidth \hbox to
\textwidth{\hss(a)\kern0em\hss(b)\kern4em} \kern+0.315\textwidth
\centerline{
\includegraphics[scale=0.35]{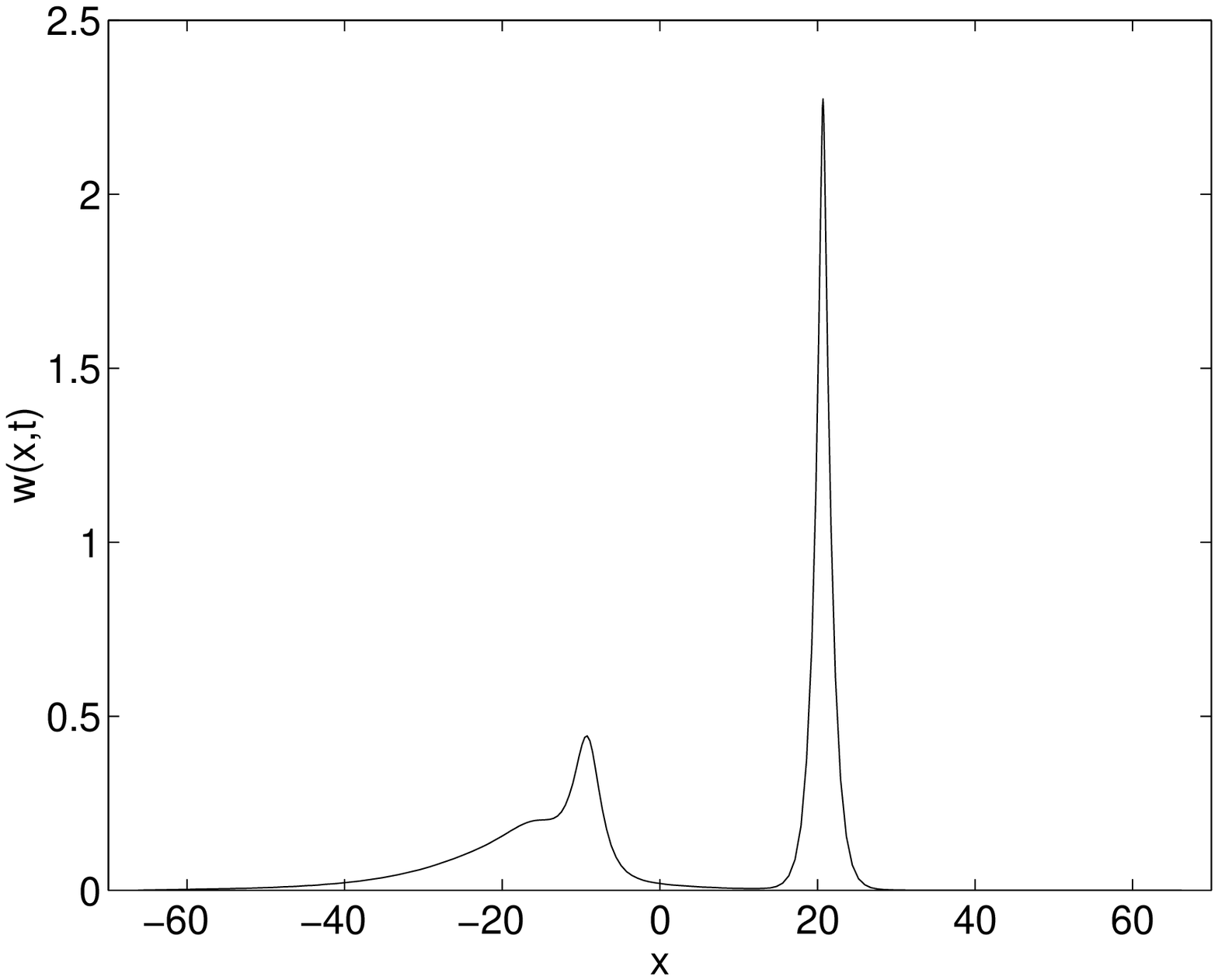}\quad
\includegraphics[scale=0.35]{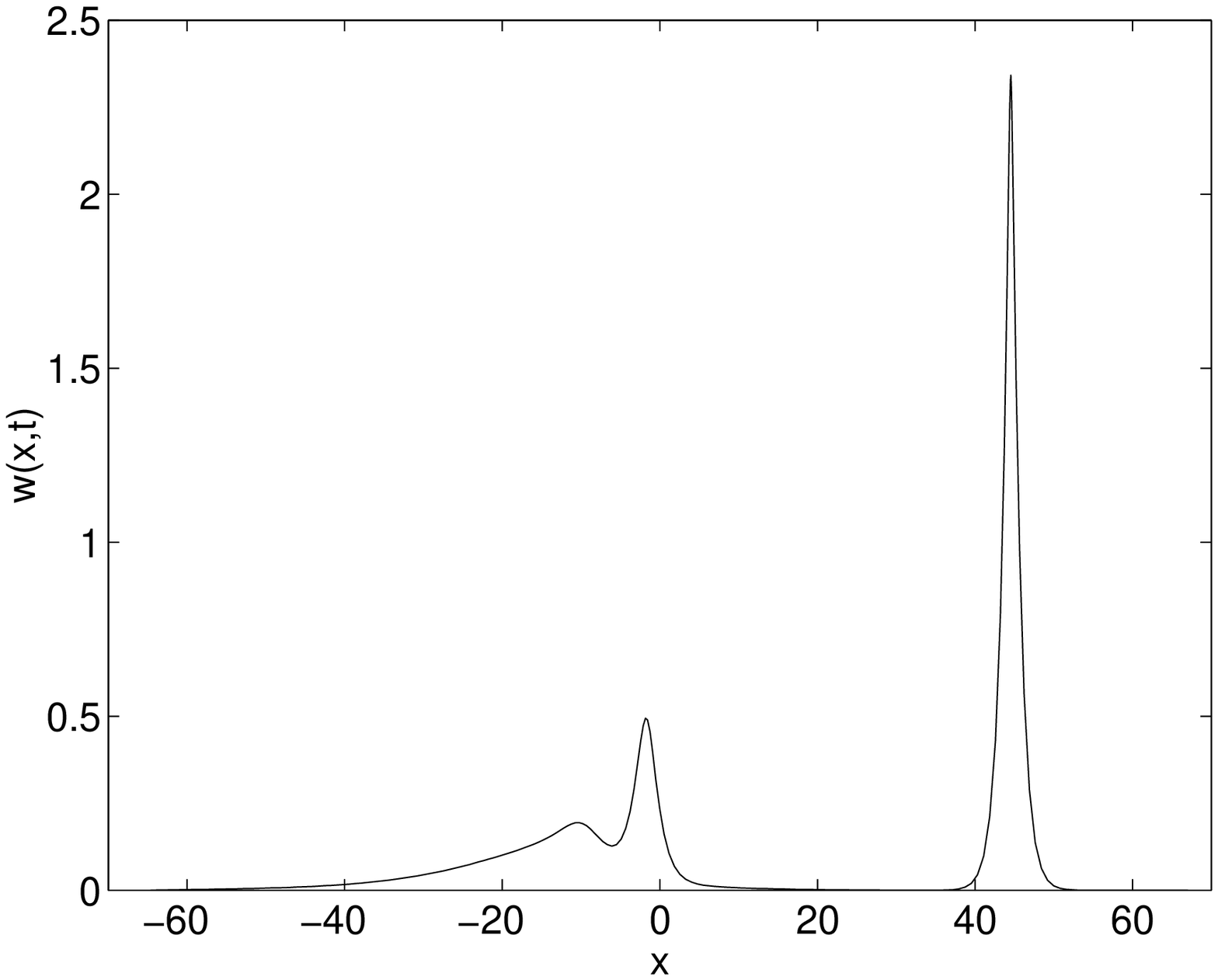}}
\kern-0.315\textwidth \hbox to
\textwidth{\hss(c)\kern0em\hss(d)\kern4em} \kern+0.315\textwidth
\caption{Numerical solution starting from an initial condition: (a)
$t=0.0$; (b) $t=10.0$; (c)
 $t=20.0$; (d) $t=30.0$.}
\label{f:nonexact}
\end{figure}

{\bf Example 4: Initial condition of non exact soliton
solutions.} We show that the integrable scheme can be
also applied for the initial value problem starting with a non-exact
solution. We choose an initial condition
in the following procedure.
The mesh
size is determined by
\begin{equation}\label{nonexact}
%%    \delta_k=2ch(1-0.8 {\rm sech}(2kh-W_x/3)),
%%%\delta_k=2{\rm cosh}(1-0.8 {\rm sech}(2kh-W_x/3)),
\delta_k=2ac(1-0.8 {\rm sech}(2ka-W_x/3))\,,
\end{equation}
where $W_x$ ($=8$) is the width of computation, $N=201$ is
the number of grid in $x$-domain,
$k=1, \cdots, N-1$, and $a=W_x/(N-1)=0.04$.
Then the initial profile can be calculated through the second equation
of the semi-discretization (\ref{semi-d-ch}).
The initial profile is plotted in Fig. 5(a). Figs. 5(b), (c) and (d)
show the evolution at $t=10,\,20,\,30$, respectively.
It can be seen that a soliton with large amplitude is
developed first, and moving fast to the right. By $t=30$, a second soliton
with small amplitude is developed and a third soliton is born from
the second soliton.

\section{Concluding remarks}

An integrable semi-discretization of the CH equation has been
presented in this paper. Determinant formulas of the 
$N$-soliton solutions 
of both the continuous and semi-discrete CH equations have
been derived. Multi-soliton, multi-cuspon and multi-soliton-cuspon 
solutions can be generated from above determinant formulas. As
further topics, we attempt to construct integrable
semi-discretizations of the Degasperis-Procesi equation and other
soliton equations possessing non-smooth solutions such as peakon,
cuspon, or loop-soliton solutions. We will address these issues in
forthcoming papers.

Applying integrable discretizations of soliton equations to
numerical computations remains a promising but not thoroughly
explored subject. In the present paper, even for relatively large
mesh sizes, very accurate numerical solutions of cuspon-cuspon and
soliton-cuspon interactions for the CH equations are achieved
through our proposed integrable semi-discrete scheme. In addition, a
numerical computation starting with an non-exact initial condition
is performed with a satisfactory result. 
It is worth noting that the integrable semi-discrete scheme of the CH 
equation is also a self-adaptive method, which is of great interest in 
the area of numerical partial differential equations. 
We intend to extend this new idea of self-adaptive method to other 
PDEs in the near future.

%%%%%%%%%%%%%%%%%%%%%%%%%%%%%%%%%%%%%%%%%%%%%%%%%%%%%%%%%%%%%%%%%%%%%%%%%
\section{Appendix}
In this appendix, we prove the differential formulas for $\tau_n$,
(\ref{x-dif})-(\ref{ys-dif}).
Let us introduce a simplified notation,
\begin{eqnarray}
\fl |{\psi^{(n_1)}},
  {\psi^{(n_2)}},
  \cdots,
  {\psi^{(n_N)}}|
 = \left|\matrix{
  \psi_1^{(n_1)} &\psi_1^{(n_2)}&\cdots
   &\psi_1^{(n_N)}\cr
  \psi_2^{(n_1)} &\psi_2^{(n_2)}&\cdots
   &\psi_2^{(n_N)} \cr
  \vdots                     &\vdots                     &
   &\vdots                     \cr
  \psi_N^{(n_1)} &\psi_N^{(n_2)} &\cdots
   &\psi_N^{(n_N)} \cr}
 \right|.
\end{eqnarray}
In this notation, we have
$\tau_n=|\psi^{(n)},\psi^{(n+1)},\cdots,\psi^{(n+N-1)}|$,
thus differentiating $\tau_n$ by $x$ and using (\ref{x-dispersion})
we obtain
\begin{eqnarray*}
&&\fl
\partial_x\tau_n=\sum_{j=0}^{N-1}
|\psi^{(n)},\psi^{(n+1)},\cdots,
\partial_x\psi^{(n+j)},\cdots,\psi^{(n+N-1)}|\\
&&\fl \qquad =\sum_{j=0}^{N-1}
|\psi^{(n)},\psi^{(n+1)},\cdots,
\psi^{(n+j+1)}+c\psi^{(n+j)},\cdots,\psi^{(n+N-1)}|\\
&&\fl \qquad =\sum_{j=0}^{N-1}
|\psi^{(n)},\psi^{(n+1)},\cdots,\psi^{(n+j+1)},\cdots,\psi^{(n+N-1)}|\\
&&\fl \qquad +\sum_{j=0}^{N-1}
|\psi^{(n)},\psi^{(n+1)},\cdots,c\psi^{(n+j)},\cdots,\psi^{(n+N-1)}|\\
&&\fl \qquad =|\psi^{(n)},\psi^{(n+1)},\cdots,\psi^{(n+N-2)},\psi^{(n+N)}|
+Nc|\psi^{(n)},\psi^{(n+1)},\cdots,\psi^{(n+N-1)}|\,,
\end{eqnarray*}
which gives (\ref{x-dif}).
Similarly differentiating $\tau_n$ by $y$ and using (\ref{y-dispersion})
we get
\begin{eqnarray*}
&&\fl
\partial_y\tau_n=\sum_{j=0}^{N-1}
|\psi^{(n)},\psi^{(n+1)},\cdots,
\partial_y\psi^{(n+j)},\cdots,\psi^{(n+N-1)}|\\
&&\fl \qquad =\sum_{j=0}^{N-1}
|\psi^{(n)},\psi^{(n+1)},\cdots,
\psi^{(n+j+2)}+2c\psi^{(n+j+1)}+c^2\psi^{(n+j)},\cdots,\psi^{(n+N-1)}|\\
&&\fl \qquad =|\psi^{(n)},\psi^{(n+1)},\cdots,\psi^{(n+N-2)},\psi^{(n+N+1)}|
-|\psi^{(n)},\psi^{(n+1)},\cdots,\psi^{(n+N-3)},\psi^{(n+N-1)},\psi^{(n+N)}|
\nonumber\\
&&\fl \qquad +2c|\psi^{(n)},\psi^{(n+1)},\cdots,\psi^{(n+N-2)},\psi^{(n+N)}|
+Nc^2|\psi^{(n)},\psi^{(n+1)},\cdots,\psi^{(n+N-1)}|\,.
\end{eqnarray*}
We obtain (\ref{y-dif}) from the above equation and (\ref{x-dif}).
Eqs.~(\ref{t-dif}) and (\ref{s-dif}) can be proved by using
(\ref{t-linear}) and (\ref{s-linear}) in a similar way.
Finally eqs.~(\ref{xt-dif})-(\ref{ys-dif}) can be verified by
differentiating (\ref{x-dif}) and (\ref{y-dif}) by $t$ and $s$
through similar calculations.

%%%%%%%%%%%%%%%%%%%%%%%%%%%%%%%%%%%%%%%%%%%%%%%%%%%%%%%%%%%%%%%%%%%%%%%%%
%\section{[***]}

%%%%%%%%%%%%%%%%%%%%%%%%%%%%%%%%%%%%%%%%%%%%%%%%%%%%%%%%%%%%%%%%%%%%%%%%%
%\section{Conclusions}

%%%%%%%%%%%%%%%%%%%%%%%%%%%%%%%%%%%%%%%%%%%%%%%%%%%%%%%%%%%%%%%%%%%%%%%%%
%\section*{Acknowledgements}

%%\newpage
%%%%%%%%%%%%%%%%%%%%%%%%%%%%%%%%%%%%%%%%%%%%%%%%%%%%%%%%%%%%%%%%%%%%%%%%%

\end{document}